\documentclass[jcp,showpacs,showkeys,preprintnumbers,amsmath,amssymb,superscriptaddress,twocolumn]{revtex4-2}
\pdfoutput=1 
\usepackage{graphicx}
\usepackage{soul}
\usepackage{amsfonts}
\usepackage{url}
\usepackage{epstopdf}
\usepackage{color}
\usepackage{tikz-cd}
\usepackage[colorlinks= true, linkcolor=blue, citecolor=blue, urlcolor=blue]{hyperref}
\usepackage{bm}
\usepackage{lipsum}
\usepackage{longtable}

\def\R{{\mathbb R}}

\def\ctanh{\mathrm{ctanh}}

\def\x{\bm{x}}

\def\Res{\mathrm{Res}}

\begin{document}

\title{Escape from textured adsorbing surfaces}

\author{Yuval Scher}
\email{yuvalscher@mail.tau.ac.il}
\affiliation{
School of Chemistry, Center for the Physics \& Chemistry of Living Systems, Ratner Institute for Single Molecule Chemistry,
and the Sackler Center for Computational Molecular \& Materials Science, Tel Aviv University, 6997801 Tel Aviv, Israel}

\author{Shlomi Reuveni}
\email{shlomire@tauex.tau.ac.il}
\affiliation{
School of Chemistry, Center for the Physics \& Chemistry of Living Systems, Ratner Institute for Single Molecule Chemistry,
and the Sackler Center for Computational Molecular \& Materials Science, Tel Aviv University, 6997801 Tel Aviv, Israel}

\author{Denis~S.~Grebenkov}
 \email{denis.grebenkov@polytechnique.edu}
\affiliation{
Laboratoire de Physique de la Mati\`{e}re Condens\'{e}e, \\ 
CNRS -- Ecole Polytechnique, IP Paris, 91120 Palaiseau, France}

\date{\today}

\begin{abstract}
\end{abstract}

\begin{abstract}
The escape dynamics of sticky particles from textured surfaces is poorly understood despite importance to various scientific and technological domains. In this work, we address this challenge by investigating the escape time of adsorbates from prevalent surface topographies, including holes/pits, pillars, and grooves. Analytical expressions for the probability density function and the mean of the escape time are derived. A particularly interesting scenario is that of very deep and narrow confining spaces within the surface. In this case, the joint effect of the entrapment and stickiness prolongs the escape time, resulting in an effective desorption rate that is dramatically lower than that of the untextured surface. This rate is shown to abide a universal scaling law, which couples the equilibrium constants of adsorption with the relevant confining length scales. While our results are analytical and exact, we also present an approximation for deep and narrow cavities based on an effective description of one-dimensional diffusion that is punctuated by motionless adsorption events. This simple and physically motivated approximation provides high-accuracy predictions within its range of validity and works relatively well even for cavities of intermediate depth. All theoretical results are corroborated with extensive Monte-Carlo simulations.
\end{abstract}

\maketitle

\section{Introduction}\label{sec:intro}

Fabrication of nanoscale surface topographies have seen rapid developments in the last two decades \cite{lord2010influence,alhmoud2021maceing,flemming1999effects}. In particular, controlled fabrication of nano-arrays can be achieved, where common structured surface topographies include nano-arrays of pillars \cite{hochbaum2005controlled,elnathan2014engineering, sun2014large,checco2014collapse,chen2022nanoforest}, arrays of holes/pits \cite{richert2010adsorption,wilkinson2011biomimetic}, and grooves \cite{tsai2009fibronectin,de2015effect,van2008manufacturing}. Alternatively, the surface can be rugged and possess random roughness \cite{coppens1999effect}. 

The effect of surface topography has proved to be a key aspect when considering heterogeneous catalysis \cite{ben2000diffusion,coppens1999effect} and the passivisation of catalytic surfaces \cite{filoche2005deactivation,filoche2008passivation}. It also plays a cardinal role when considering living cell behavior, as protein adsorption to a textured surface mediates the cell attachment to the surface \cite{lord2010influence,flemming1999effects,chen2009effect}. Topographical features can affect the adsorption properties of a protein by inducing conformational changes, or by other forms of surface-protein interactions \cite{roach2007modern}. When the length scale of the topographical features is larger than the protein size, additional effects come into play. Often, adsorption is increased as a larger number of active sites for protein adsorption are available. Another crucial effect is the entrapment of the proteins inside confined spaces \cite{wilson2005mediation,peng2013micropatterned}.  

The entrapment effect was vividly illustrated in a series of on-chip devices made by the Patolsky group \cite{borberg2019light,borberg2021depletion,krivitsky2012si}. These devices utilized entrapment in sticky confined spaces of textured surfaces for the purpose of selective separation of required protein analytes from raw biosamples. The selective stickiness was achieved by attaching specific antibodies to the surface. The surface was textured by a vertical array of nanopillars, albeit other topographies, like grooves, are expected to exhibit similar behavior. The Patolsky group demonstrated that the target proteins are entraped in the surface for extremely long times (weeks and even months). This came as a surprise, since the same antibody, if used on a flat surface, would bind the biomolecules for a few milliseconds only. Similarly, using the nanopillar vertical array without antibodies leads to fast diffusive escape. The dramatic effect of prolonged escape times is hence due to a combination of topography and adsorption/stickiness. A semi-quantitative explanation of the experimental results was given in Ref. \cite{borberg2019light}. 

Qualitatively, when the confining space is deep and narrow, the escaping particle is forced to collide with the confining walls a large number of times before it can escape. Each collision can result in an adsorption event, and these add up and eventually culminate in extremely prolonged escape. Thus, despite the relatively short dissociation time from the antibody, and due to the multitude of adsorption events, the textured surface appears as if it has a very high affinity to the protein. This observation is an important first step, but a more detailed quantitative understanding is currently missing. The challenge is thus to determine how exactly does surface topography and texture affect the escape from a surface. Specifically, how does the mean escape time scale with the depth and width of confining spaces within the surface? and how does sticky entrapment affect the statistics of the escape time, as characterized by its probability density function (PDF)? 

Recently, we have developed an analytical approach that allows one to provide an exact solution to the aforementioned problems \cite{scher2023escape}. We considered the escape of a diffusing particle from a domain of arbitrary shape, size, and surface reactivity. The escape time from the adsorbing confining spaces of a textured surface can be computed using this formalism. Here, we perform this calculation for three different topographies of adsorbing surfaces: (i) a surface perforated with pits/holes is considered in Sec. \ref{sec:holes}; (ii) a surface textured by an array of pillars is considered in Sec. \ref{sec:pillars}; (iii) a surface textured by grooves is considered in Sec. \ref{sec:grooves}. 

For each of the above-mentioned cases, we aim to find the escape time of a particle initially entrapped in the confining spaces of the textured surface. We assume that the surface is homogeneously textured, i.e., all of the confining spaces are of the same size and repeated periodically. Thus, the escape time out of the periodic cell in fact equals to the escape time from the textured surface. Note that in some cases, e.g., a surface perforated with holes, the assumption of periodicity can be easily relaxed -- it is the homogeneity which is important. Lastly, while in this work we consider homogeneously textured surfaces, the same formalism can be used when dealing with heterogeneous textured surfaces with a known size distribution of the confining spaces: The escape time from the textured surface will be the appropriately weighted sum of the escape times from cells of different sizes. 
 
Alongside exact results, we also present an insightful approximation. In Sec. \ref{two_state_diffsion} we introduce a two-state switching diffusion approximation for the diffusive escape from sticky nanocavities. This approximation is appropriate for deep and narrow cavities, where diffusion is effectively one-dimensional. The adsorption to the surface is then effectively accounted for by the introduction of an immobile state for which the diffusion coefficient vanishes. We illustrate that this approximation is very accurate, and works well even for cavities of intermediate depth. We utilize this approximation yet again in Sec. \ref{sec:grooves_decay}, where we calculate the asymptotic decay rate of the escape time PDF. Indeed, the approximation is expected to work in the limit of very deep cavities regardless of the lateral geometry. Thus, a central benefit of the two-state switching diffusion approximation is that it captures the essential physics of the problem at hand and simplifies the analysis without losing much in accuracy. 

The three problems solved here abide similar laws and show similar characteristic behavior. In Sec. \ref{sec:discussion} we discuss a general form of the equation for the mean escape time and for its inverse, which is the effective desorption rate from the textured surface. This suggests that the results presented here are universal in nature and can be applied, even if approximately, when considering more complicated scenarios.


\begin{figure}[!h]
\begin{center}
\includegraphics[width=70mm]{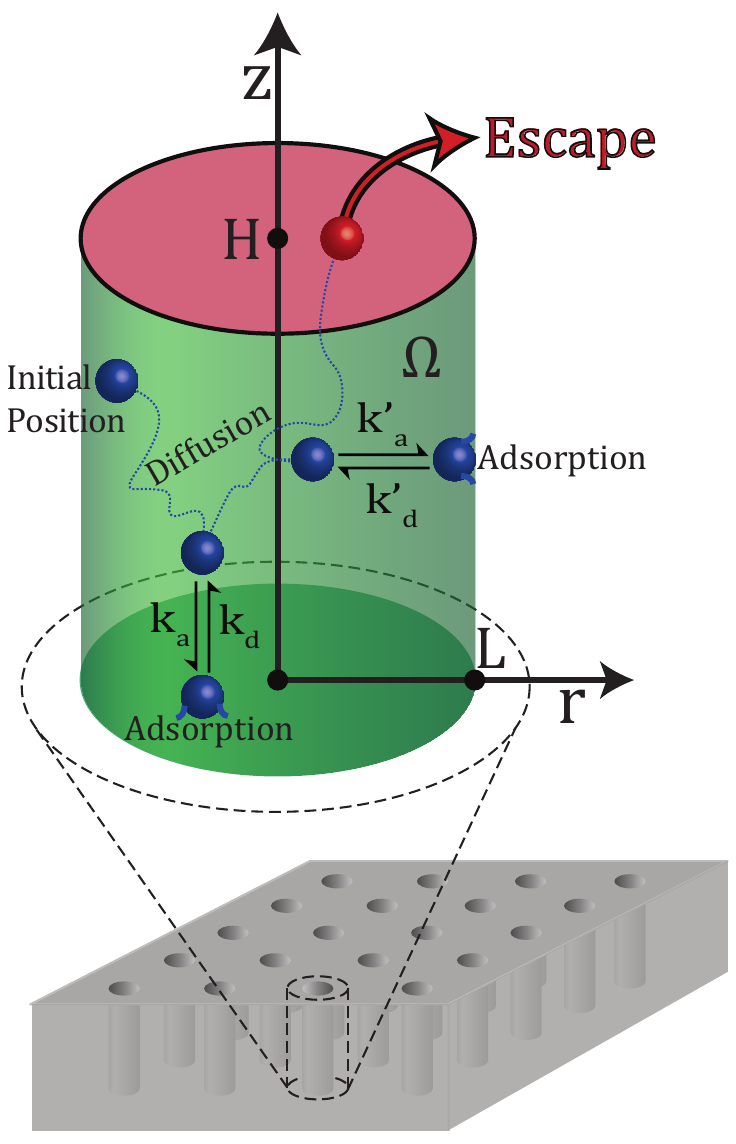}
\end{center}
\caption{
A perforated surface with cylindrical holes. One of the holes is enlarged: A cylinder of radius $L$ capped by parallel planes at $z = 0$ and $z =H$.  The top disk at $z = H$ is absorbing (escape region in red), whereas the bottom disk at $z = 0$ and the cylindrical wall are adsorbing (green), with reversible binding kinetics. Here, $k_a$ and $k_d$ are the adsorption and desorption constants for the bottom disk, and $k'_a$ and $k'_d$ are the adsorption and desorption
constants for the cylindrical surface.}
\label{fig:scheme_3D}
\end{figure}

\section{Adsorbing perforated surface}\label{sec:holes}

We consider a perforated surface with cylindrical holes, illustrated in Fig. \ref{fig:scheme_3D}. We aim to find the escape time of a particle initially entrapped inside one hole of the surface. Our task is thus calculating the escape time from the periodic cell.  For the example under consideration here, the periodic cell is a cylindrical hole. A representative hole is enlarged in Fig. \ref{fig:scheme_3D}. Escaping from the textured surface is thus equivalent to the escape problem in three dimensions when the particle
diffuses inside a cylinder of radius $L$ capped by parallel planes at
$z = 0$ and $z = H$.  The top disk is
absorbing whereas the remaining boundary of the domain is {\it
adsorbing}. We assume the adsorption kinetics is linear and homogeneous on each surface, but we allow for heterogeneity in the sense that the adsorption and desorption rates on the bottom disk and on the curved cylindrical surface can differ. We are interested in finding the PDF of the first-passage time to the top disk, which can also be thought of as the escape time from the cylindrical compartment. We denote this PDF as $J_{\rm{ab}}(t|r,z)$, where $(r,z) \in \Omega$ is the initial location of the particle inside the cylindrical domain $\Omega$. In Ref. \cite{scher2023escape} we have derived, for the general case, the partial differential equation and boundary conditions governing the Laplace transform of this PDF, $\tilde{J}_{\rm{ab}}(s|r,z) = \int\nolimits_0^{\infty} dt\,  e^{-ts} J_{\rm{ab}}(t|r,z)$. For the specific geometry considered here, 
these equations simplify to
\begin{subequations}
\begin{align} \label{differential_equation_a}
(s - D\Delta) \tilde{J}_{\rm{ab}}(s|r,z) & = 0 \quad (r,z \in \Omega), \\ \label{differential_equation_b}
\tilde{J}_{\rm{ab}}(s|r,z) & = 1 \quad (z = H) ,\\ \label{differential_equation_c}
(-\partial_{z} + q_s) \tilde{J}_{\rm{ab}}(s|r,z) & = 0 \quad (z = 0), \\ \label{differential_equation_d}
(\partial_r + q'_s) \tilde{J}_{\rm{ab}}(s|r,z) & = 0 \quad (r = L), 
\end{align}
\end{subequations}
where $\Delta = \partial_{r}^2 + (1/r)\partial_{r} + \partial_{z}^2$ is the
Laplace operator in cylindrical coordinates (without the angular part), and $D$ is the diffusion coefficient. The surfaces are characterized by the parameters $q_s$ and $q'_s$:
\begin{equation} \label{eq:qs}
q_s = \frac{k_a}{D(1 + k_d/s)} \,, \qquad q'_s = \frac{k'_a}{D(1 + k'_d/s)} \,,
\end{equation}
where $k_a$ and $k_d$ are the adsorption and desorption constants for the
bottom disk, and $k'_a$ and $k'_d$ are the adsorption and desorption
constants for the cylindrical surface.

\subsection{Solution in Laplace domain}\label{holes_Laplace}

We search the solution of Eq. (\ref{differential_equation_a}) under the boundary conditions (\ref{differential_equation_b})-(\ref{differential_equation_d}) as 
\begin{equation}  \label{eq:tildeJ_general}
\tilde{J}_{\rm{ab}}(s|r,z) = 2 \sum\limits_{n=0}^\infty c_n^{(s)} J_0(\alpha_n^{(s)} \bar{r}) \frac{g_n^{(s)}(z)}{g_n^{(s)}(H)} \,,
\end{equation}
where $\bar{r}=r/L$, $J_\nu(\cdot)$ is the Bessel function of the first kind of order $\nu$ and
\begin{equation} \label{eq:g_def}
g_n^{(s)}(z) = \hat{\alpha}_n^{(s)} \cosh(\hat{\alpha}_n^{(s)} \bar{z}) + q_s L \sinh(\hat{\alpha}_n^{(s)} \bar{z})
\end{equation}
to respect the boundary condition (\ref{differential_equation_c}), with $\bar{z}=z/L$
and 
\begin{equation} \label{eq:hat_alpha}
\hat{\alpha}_n^{(s)} = \sqrt{[\alpha_n^{(s)}]^2 + L^2 s/D} \,.
\end{equation}
From the boundary condition (\ref{differential_equation_d}) we find that
$\alpha_n^{(s)}$ satisfy the transcendental equation
\begin{equation}  \label{eq:alphan_eq}
\alpha_n^{(s)} \frac{J_1(\alpha_n^{(s)})}{J_0(\alpha_n^{(s)})} = q'_s L.
\end{equation}
For any $s\geq 0$, there are infinitely many solutions of this equation that we enumerate by $n = 0,1,2,\ldots$ in an increasing order.
The unknown
coefficients $c_n^{(s)}$ are found by multiplying the boundary condition (\ref{differential_equation_b}) by $rJ_0(\alpha_k^{(s)} r/L)$ and
integrating over $r$ from $0$ to $L$. This gives
\begin{align}
2c_k^{(s)} L^2 \frac{J_0^2(\alpha_k^{(s)}) + J_1^2(\alpha_k^{(s)})}{2} 
&=\underbrace{\int\limits_0^L dr \, r \, J_0(\alpha_k^{(s)} \bar{r})}_{=L^2 J_1(\alpha_k^{(s)})/\alpha_k^{(s)}} \,, 
\end{align}
from which we get
\begin{equation} \label{c_coef}
c_n^{(s)} = \frac{J_1(\alpha_n^{(s)})/\alpha_n^{(s)}}{J_0^2(\alpha_n^{(s)}) + J_1^2(\alpha_n^{(s)})} \,.
\end{equation}
To obtain these relations we used the orthogonality of the Bessel functions together with
\begin{equation}
\int\limits_0^1 dr \, r \, J_0^2(\alpha r) = \frac{J_0^2(\alpha) + J_1^2(\alpha)}{2}
\end{equation}
and
\begin{equation}
\int\limits_0^1 dr \, r \, J_0(\alpha r) = \frac{J_1(\alpha)}{\alpha} \,.
\end{equation}

It is worth noting that the numerical inversion of the Laplace
transform is challenging here; in fact, one needs to evaluate the
solution at complex $s$, which in turn requires an improved algorithm for
finding the roots $\alpha_n^{(s)}$, given that $q_s$ and $q'_s$ become
complex as well.

To facilitate further analysis, we introduce the following
dimensionless quantities:
\begin{equation} \label{further_notation}
\kappa_a = \frac{k_a L}{D} , \hspace{1.6ex} \kappa_d = \frac{k_d L^2}{D} , \hspace{1.6ex} \kappa'_a = \frac{k'_a L}{D} , \hspace{1.6ex} \kappa'_d = \frac{k'_d L^2}{D} .
\end{equation}
For convenience, in Table \ref{long} we collect the definitions of all dimensionless quantities defined so far, and also define $h=H/L$ and $\rho=l/L$ that will be used later.

\begin{table}[t!]
 \begin{tabular}{|c|} \hline
 $h=H/L$\\\hline
 $\rho=l/L$\\\hline
 $\bar{z}=z/L$\\\hline
 $\bar{r}=r/L$\\\hline
 $\kappa_a = k_a L/D$\\\hline
 $\kappa_d = k_d L^2 / D$\\\hline
 $\kappa'_a = k'_a L/D$\\\hline
 $\kappa'_d = k'_d L^2 / D$\\\hline
 $\hat{\alpha}_n^{(s)} = \sqrt{[\alpha_n^{(s)}]^2 + L^2 s/D}$\\\hline
\end{tabular}
\caption{Summary of dimensionless quantities.}
\label{long}
\end{table}

\subsection{Mean Escape Time}\label{holes_mean}

In this section, we compute the mean escape time by studying the asymptotic
behavior of $\tilde{J}_{\rm{ab}}(s|r,z)$ as $s\to 0$.  In the spectral expansion
(\ref{eq:tildeJ_general}), we first analyze the term $n = 0$ and then
discuss the other terms with $n > 0$.

As $s\to 0$, one has $q'_s\to 0$ so that $\alpha_0^{(s)}\to 0$.  Using
the Taylor series expansion of Bessel functions in
Eq. (\ref{eq:alphan_eq}), one gets in the leading order
\begin{equation}
\alpha_0^{(s)} \approx \sqrt{2q'_s L} \approx \sqrt{s} \sqrt{2\kappa'_a/k'_d} \quad (s\to 0).
\end{equation}
As a consequence, one has
\begin{equation}
c_0^{(s)} = \frac12 \biggl(1 + \frac{[\alpha_0^{(s)}]^2}{8} + \ldots\biggr) = \frac12 + \frac{\kappa'_a}{8k'_d} s + O(s^2).
\end{equation}
and
\begin{equation}
J_0(\alpha_0^{(s)} r/L) = 1 - \frac{r^2 \kappa'_a}{2L^2 k'_d} s + O(s^2).
\end{equation}
We also get $\hat{\alpha}_0^{(s)} \approx L \sqrt{s/D} \sqrt{1 +
2\kappa'_a/\kappa'_d}$ in the leading order, from which
\begin{align}
&\frac{g_0^{(s)}(z)}{g_0^{(s)}(H)}  = \\
& \hspace{5ex} 1 - \biggl(\frac{\kappa_a}{k_d} \frac{H-z}{L} + \biggl[\frac{L^2}{2D} + \frac{\kappa'_a}{k'_d}\biggr]
\frac{H^2-z^2}{L^2}\biggr) s + O(s^2). \nonumber
\end{align}

Let us now consider the terms with $n > 0$.  Denoting the left-hand
side of Eq. (\ref{eq:alphan_eq}) as $F_1(z) = zJ_1(z)/J_0(z)$, we apply
the Taylor expansion near $\alpha_n^{(0)} > 0$:
\begin{equation} \label{F_taylor}
F_1(\alpha_n^{(s)}) \approx F_1(\alpha_n^{(0)}) + F_1'(\alpha_n^{(0)}) (\alpha_n^{(s)} - \alpha_n^{(0)}) .
\end{equation}
According to Eq. (\ref{eq:alphan_eq}), $F_1(\alpha_n^{(s)}) = q'_s L \to
0$ as $s \to 0$ and thus $\alpha_n^{(0)} =
j_{1,n}$, where $j_{1,n}$ denote the zeros of the Bessel function $J_1(z)$.
Plugging in $z =
j_{1,n}$ into the relation $F_1'(z) = z(1 + J_1^2(z)/J_0^2(z))$, we find that $F_1'(j_{1,n}) = j_{1,n}$. All that remains is to compare the right-hand side of Eqs. (\ref{F_taylor}) and (\ref{eq:alphan_eq}) in the limit $s \to 0$, which gives 
\begin{equation} \label{alpha_n_taylor}
\alpha_n^{(s)} = j_{1,n} + \frac{\kappa'_a}{k'_d j_{1,n}} s + O(s^2).
\end{equation}
Similarly, setting $F_2(z) = \left(J_1(z)/z\right)/\left(J_0^2(z) + J_1^2(z)\right)$ such that according to  Eq. (\ref{c_coef}) we have $c_n^{(s)}  = F_2(\alpha_n^{(s)})$, and using Eq. (\ref{alpha_n_taylor}), we obtain
\begin{align} \nonumber
c_n^{(s)} & = F_2(\alpha_n^{(0)}) + F'_2(\alpha_n^{(0)}) (\alpha_n^{(s)} - \alpha_n^{(0)}) + O(s^2) \\
& = \frac{1}{j_{1,n} J_0(j_{1,n})} \, \frac{\kappa'_a}{k'_d j_{1,n}} s + O(s^2),
\end{align}
and
\begin{equation}
\hat{\alpha}_n^{(s)} = j_{1,n} + \frac{1+2\kappa'_a/\kappa'_d}{2j_{1,n}}\, \frac{L^2 }{D} s + O(s^2).
\end{equation}
Since $c_n^{(s)} \propto s$ for $n > 0$, the other factors in Eq. (\ref{eq:tildeJ_general}) can be found to the lowest order in $s$. This gives
\begin{equation}
\frac{g_n^{(s)}(z)}{g_n^{(s)}(H)}  \approx \frac{\cosh(j_{1,n} \bar{z})}{\cosh(j_{1,n} h)},
\end{equation}
where $h=H/L$, and
\begin{equation}
J_0(\alpha_n^{(s)} \bar{r}) \approx J_0(j_{1,n} \bar{r}).
\end{equation}
Substituting these expressions into Eq. (\ref{eq:tildeJ_general}), we
get
\begin{equation}
\tilde{J}_{\rm{ab}}(s|r,z) = 1 - s\langle \mathcal{T} (r,z) \rangle + O(s^2),
\end{equation}
where
\begin{align} \nonumber  \label{eq:Tmean}
\langle \mathcal{T} (r,z) \rangle & = \frac{H^2-z^2}{2D} + \frac{k_a(H-z)}{k_d D}  \\  \nonumber
& + \frac{k'_a L}{k'_d D} \biggl(\frac{H^2 - z^2}{L^2} + \frac{2r^2 - L^2}{4L^2} \\  
& - 2 \sum\limits_{n=1}^\infty \frac{J_0(j_{1,n} \bar{r})}{j_{1,n}^2 J_0(j_{1,n})} \,  
\frac{\cosh(j_{1,n} \bar{z})}{\cosh(j_{1,n} h)} \biggr)
\end{align}
is the mean escape time.  

In the limit $k'_a \to 0$ or $k'_d \to \infty$, namely when the cylindrical surface is not sticky but reflecting, we retrieve the mean escape time for a one-dimensional box with a sticky surface \cite{scher2023escape} 
\begin{equation}\label{24}
\langle \mathcal{T} (r,z) \rangle =
\frac{1}{2D}(H^2-z^2) + \frac{k_a}{k_d D}(H-z).
\end{equation}

The average of
Eq. (\ref{eq:Tmean}) over the cross-section at a fixed height $z$ yields
\begin{align} \nonumber
\langle\overline{\mathcal{T}}(z)\rangle & = \frac{2\pi}{\pi L^2} \int\limits_0^L dr \, r\, \langle \mathcal{T} (r,z) \rangle \\    \label{eq:Tmean_average}
& = \frac{k_a}{k_d}  \frac{H-z}{D} + \biggl(1 + \frac{2k'_a}{k'_d L}\biggr) \frac{H^2-z^2}{2D}  ,
\end{align}
where we used $\int_0^L dr J_0(j_{1,n} r/L) = 0$. This remarkably simple expression quantifies the effect of
adsorption/desorption mechanisms onto the mean escape time. Further averaging over $z$, we obtain 
\begin{equation} \label{mean_uniform}
\langle \mathcal{T}_{u}\rangle  = \frac{H^2}{3D} \left(1 +  \frac{k_a}{k_d} \frac{3}{2H}  + \frac{k'_a}{k'_d}\frac{2}{L}  \right),
\end{equation}
where the subscript `$u$' denotes a uniform distribution of the initial position within the cylinder, which is a common experimental condition. Indeed, zooming out and assuming that the particle's initial position is distributed uniformly inside the holes of the textured surface, it is equally likely to find the particle in any of the holes. Since the surface is homogeneous and the holes are all the same, the mean escape time from the textured surface is thus given by Eq. (\ref{mean_uniform}). 

\begin{figure*}[t]
\begin{center}
\includegraphics[width=\linewidth]{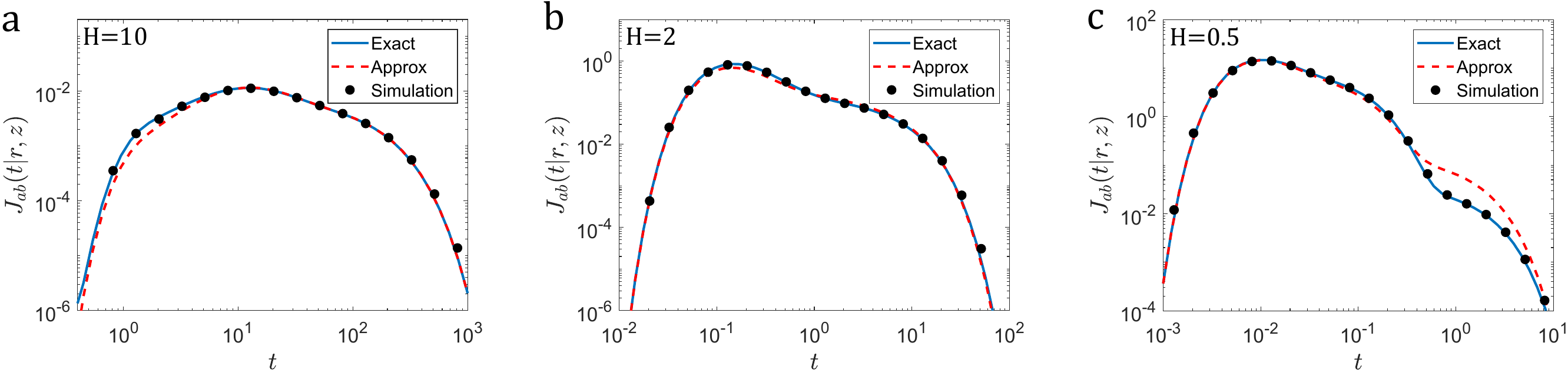}
\end{center}
\caption{PDF $J_{\rm ab}(t|r,z)$ of the escape time from a cylindrical hole (see Fig. \ref{fig:scheme_3D}) with $L = 1$, $k_a = 0$, $D = 1$, $k'_a = 1$, $k'_d = 1$, $r = 0$, $z = H/2$, and with $H = 10$ (left), $H = 2$ (center), and $H = 0.5$ (right).  Solid lines represent the exact solution from Eq. (\ref{eq:Jt_3D}) truncated to $30 \times 30 = 900$ terms.  Dashed lines give the two-state switching diffusion approximation $J_{\rm sd}(t|z)$ from Eq. (\ref{eq:Jt_two}) truncated to 200 terms. Circles give estimates based on $10^6$ particles whose motion was simulated according to the the protocol in Appendix D of Ref. \cite{scher2023escape}, with simulation time step $\Delta t=10^{-6}$. }
\label{fig:3D_Jt}
\end{figure*}

\subsection{Two-state switching diffusion approximation}\label{two_state_diffsion}

In Sec. \ref{holes_time_domain} we will analytically invert Eq. (\ref{eq:tildeJ_general}) to get the PDF of the escape time. As this inversion is quite involved, it is worthwhile to first consider a simple approximation for the problem at hand: a model of two-state switching diffusion. This model is expected to approximate the escape from a perforated surface in the limit $L \ll H$, i.e., when the holes are very narrow and deep. In this limit, assuming that the reactivities of all surfaces are comparable, the area of the lower disk becomes negligible compared to the area of the cylindrical surface. The bottom disk is thus expected to have little effect on the escape time and so we can treat it as an inert reflecting surface ($k_a = 0$).  

When $L\ll H$, one can expect that diffusion in the radial direction is not  relevant and try to reduce the original model to a two-state switching diffusion model when the particle diffuses in the bulk with a diffusion coefficient $D$ in the state 1, or remains immobile in the state 2 (mimicking its adsorbed state). The transition between two states is a random first-order kinetics with rates $k_{12}$ from the free state to the adsorbed state and $k_{21} = k'_d$ in the reverse direction (see below). A model of two-state switching free diffusion was introduced by Kärger \cite{karger1985nmr} and more general models were studied in \cite{Godec17,Grebenkov19a} (see references therein).  In particular, the subordination concept was used in \cite{Grebenkov19a} to show that the PDF of the escape time admits a general spectral expansion 
\begin{equation}
J_{\rm sd}(t|\x) = - \sum\limits_{n=0}^\infty \partial_t \Upsilon(t;\Lambda_n) u_n(\x) \int\limits_{\Omega} d\x' \, u_n(\x'),
\end{equation}
where $\Lambda_n$ and $u_n(\x)$ are the eigenvalues and
$L_2(\Omega)$-normalized eigenfunctions of the Laplace operator with
appropriate boundary conditions, and the function
$\Upsilon(t;\Lambda)$ accounts for switching dynamics.  An explicit
form of this function  for the two-state model
\cite{karger1985nmr,Grebenkov19a} reads in our setting as
\begin{equation}
\Upsilon(t;\Lambda) = \frac{e^{-\gamma_+ t}(D\Lambda - \gamma_-) - e^{-\gamma_- t}(D\Lambda - \gamma_+)}{\gamma_+ - \gamma_-} \,,
\end{equation}
where
\begin{align} \label{gamma_def}
&\gamma_{\pm} = \\
&\frac12 \left(D\Lambda + k_{12} + k_{21} \pm \sqrt{(D\Lambda + k_{21} + k_{12})^2 - 4D\Lambda k_{21}}\right). \nonumber
\end{align}

Here we consider diffusion on the interval $(0,H)$ with an absorbing
endpoint $z = H$ and a reflecting endpoint $z = 0$, for which $\Lambda_n
= \pi^2 (n+1/2)^2/H^2$ and $u_n(z) = \sqrt{2/H} \cos(\pi (n+1/2)z/H)$, such that
\begin{align}  \label{eq:Jt_two}
&J_{\rm sd}(t|z) =
\\
& \sum\limits_{n=0}^\infty \partial_t \Upsilon(t;\Lambda_n) \frac{2(-1)^{n+1}}{\pi (n+1/2)} \cos(\pi(n+1/2)z/H). \nonumber
\end{align}
From this PDF, we calculate the mean escape time:
\begin{align} \nonumber
T_{\rm sd}(z) &= \int\limits_0^\infty dt \, t \, J(t|z) = \sum\limits_{n=0}^\infty \frac{2(-1)^n}{\pi(n+1/2)} \\  \nonumber
& \times \cos(\pi (n+1/2)z/H) \int\limits_0^\infty dt \, \Upsilon(t;\Lambda_n) \\  \nonumber
& = \frac{(k_{12} + k_{21})H^2}{Dk_{12}} \sum\limits_{n=0}^\infty \frac{2(-1)^n \cos(\pi(n+1/2)z/H)}{\pi^3 (n+1/2)^3} \\  \label{eq:Tmean_switch}
& = \bigl(1 + k_{12}/k_{21}) \frac{H^2-z^2}{2D} \,,
\end{align}
where we note that the sum in the third line is the Fourier series of $( 1 - (z/H)^2)/2$.
Comparison of Eqs. (\ref{eq:Tmean_average}) and (\ref{eq:Tmean_switch})
suggests a way to assign the transition rates as $k_{12} = 2k'_a/L$ and $k_{21}=k'_d$ so that the MFPTs are identical in both cases (for $k_a=0$). Note that for $k_a>0$, the approximation identifies with the exact result to the leading order in $H/L$ which was hereby considered large.

In Fig. \ref{fig:3D_Jt} we assume $k_a=0$ and demonstrate how the two-state diffusion approximation $J_{\rm sd}(t|z)$ captures the PDF $J_{\rm{ab}}(t|r,z)$ in the limit $L \ll H$. We plot the density and its approximation for three heights $H$ of the cylinder with unit radius $L=1$. We see that even when $H \approx L$, the two-state approximation turns out to be remarkably accurate at long times. Surprisingly, it accurately captures even the short-time behavior.

\subsection{Solution in time domain}\label{holes_time_domain}

The desorption kinetics implies the $s$-dependence of the parameters
$q_s$ and $q'_s$ in the Robin boundary condition and thus leads to a
convolution-type boundary condition in time domain, rendering the
problem much more difficult than that with the ordinary Robin boundary
condition for irreversible binding.  Nevertheless, as diffusion is
restricted in a bounded domain, the PDF of the escape time is still expected
to admit a spectral expansion.  Moreover, the presence of an absorbing
boundary at $z = H$ ensures that the survival probability vanishes
exponentially in the long-time limit.  

In mathematical terms, the inversion of the Laplace transform
$\tilde{J}_{\rm{ab}}(s|r,z)$ can be performed by evaluating its Bromwich integral representation via the residue theorem
\begin{align} \nonumber
J_{\rm{ab}}(t|r,z) &= \frac{1}{2\pi i}\int\limits_{\gamma} ds \, e^{st} \, \tilde{J}_{\rm{ab}}(s|r,z) \\
& = \sum\limits_j e^{s_j t} \, \Res_{s_j}\{ \tilde{J}_{\rm{ab}}(s|r,z) \} ,
\end{align}
where $\{s_j\}$ are the poles of $\tilde{J}_{\rm{ab}}(s|r,z)$, $\Res_{s_j}\{
\tilde{J}(s|r,z) \}$ is its residue at $s_j$, and $\gamma$ is a
contour in the complex plane of $s$ chosen such that all the poles are
to the left of it.  As the poles are determined by zeros of the function
$g_n^{(s)}(H)$ in Eq. (\ref{eq:tildeJ_general}), it is more convenient
to employ a double index $(n,m)$ instead of a single index $j$.  In
fact, the first index $n = 0,1,2,\ldots$ refers to the function
$g_n^{(s)}(H)$, whereas the second index $m = 0,1,2,\ldots$ enumerates
all positive zeros $s_{n,m}$ of this function:
\begin{equation}\label{equationforthepoles}
g_n^{(s_{n,m})}(H) = 0.
\end{equation}
The numerical computation of the poles $s_{n,m}$ will be discussed in
Sec. \ref{holes_poles}.

To compute the residues, we need to find
\begin{align} \label{eq:dgnH_ds}
& \frac{dg_n^{(s)}(H)}{ds} = \frac{d\hat{\alpha}_n^{(s)}}{ds} \biggl(\cosh(\hat{\alpha}_n^{(s)} h) 
+ h \hat{\alpha}_n^{(s)} \sinh(\hat{\alpha}_n^{(s)} h) \biggr) \nonumber \\
& + L\biggl(\frac{dq_s}{ds} \sinh(\hat{\alpha}_n^{(s)} h) + q_s \frac{d\hat{\alpha}_n^{(s)}}{ds} h \cosh(\hat{\alpha}_n^{(s)} h)\biggr),
\end{align}
where 
\begin{equation}
\frac{dq_s}{ds} = \frac{k_a k_d}{D(k_d + s)^2}, 
\end{equation}
and
\begin{equation}
\frac{d\hat{\alpha}_n^{(s)}}{ds} = \frac{1}{2\hat{\alpha}_n^{(s)}} \biggl(\frac{L^2}{D} + 2\alpha_n^{(s)} \frac{d\alpha_n^{(s)}}{ds}\biggr).
\end{equation}
If there are higher-order poles one would need to evaluate higher-order derivatives with respect to s. However, we did not observe numerically higher-order poles for all examples considered in the work. The derivative of $\alpha_n^{(s)}$ can be obtained by differentiating
Eq. (\ref{eq:alphan_eq}):
\begin{equation}\label{alpha_der}
\frac{d\alpha_n^{(s)}}{ds} \alpha_n^{(s)} \left( 1 +  \frac{J_1^2(\alpha_n^{(s)})}{J_0^2(\alpha_n^{(s)})}\right)
= \frac{k'_a k'_d L}{D(k'_d + s)^2} \,.
\end{equation}
Substituting $s_{n,m} = -D\lambda_{n,m}/L^2$, we get
\begin{align}  
& \left. \frac{dg_n^{(s)}(H)}{ds} \right|_{s=s_{n,m}}  \\ \nonumber 
& = i\biggl(h\beta_{n,m} \sin(\beta_{n,m} h) - \cos(\beta_{n,m} h) 
\biggl(1 - \frac{h\kappa_a \lambda_{n,m}}{\kappa_d - \lambda_{n,m}}\biggr)\biggr) \\  \label{eq:dgnH_ds}
& \times \frac{L^2/D + 2\alpha_{n,m} \frac{d\alpha_n^{(s_{n,m})}}{ds}}{2\beta_{n,m}} 
+ i \sin(\beta_{n,m} h) \frac{\kappa_a \kappa_d L^2/D}{(\kappa_d - \lambda_{n,m})^2} \,,\nonumber
\end{align}
where $\alpha_{n,m} = \alpha_n^{(s_{n,m})}$, $\beta_{n,m} = -i
\hat{\alpha}_n^{(s_{n,m})}$, and
\begin{equation}
 \alpha_{n,m} \left. \frac{d\alpha_n^{(s)}}{ds} \right|_{s = s_{n,m}} 
= \frac{\kappa'_a \kappa'_d L^2}{D(\kappa'_d - \lambda_{n,m})^2 \left(1 + \frac{J_1^2(\alpha_{n,m})}{J_0^2(\alpha_{n,m})}\right)} 
,
\end{equation} 
which is obtained by use of Eq. (\ref{alpha_der}). We  therefore get
\begin{align} \label{eq:Jt_3D}
&J_{\rm{ab}}(t |r,z)  = \sum\limits_{n,m=0}^\infty e^{-Dt\lambda_{n,m}/L^2} c_{n,m} J_0(\alpha_{n,m} r/L) \\  
& \times \biggl(\beta_{n,m} \cos(\beta_{n,m} z/L) + \frac{\kappa_a }{1 - \kappa_d/\lambda_{n,m}} \sin(\beta_{n,m} z/L)\biggr), \nonumber
\end{align}
where
\begin{align}  \label{eq:cnl}
c_{n,m} & = \frac{2ic_n^{(s_{n,m})}}{\frac{dg_n^{(s)}(H)}{ds}\bigr|_{s=s_{n,m}}} \\  
& = \frac{2i}{\frac{dg_n^{(s)}(H)}{ds}\bigr|_{s=s_{n,m}}} \, 
\frac{J_1(\alpha_{n,m})/\alpha_{n,m}}{J_0^2(\alpha_{n,m}) + J_1^2(\alpha_{n,m})} \,.\nonumber
\end{align}


\subsection{Poles}
\label{holes_poles}

The poles of $\tilde{J}_{\rm{ab}}(s|r,z)$ are determined by zeros of the
function $g_n^{(s)}(H)$ in Eq. (\ref{eq:tildeJ_general}).  Let us fix
$n$ and introduce the shortcut notations $\alpha_n^{(s)} = \alpha$ and
$\hat{\alpha}_n^{(s)} = i\beta$, so that
\begin{equation}  \label{eq:system1}
\alpha^2 + \beta^2 = \lambda = - sL^2/D > 0.
\end{equation}
Here $\alpha$ is the solution of Eq. (\ref{eq:alphan_eq}) that we
rewrite explicitly as
\begin{equation}  \label{eq:system2}
\alpha \frac{J_1(\alpha)}{J_0(\alpha)} = \frac{\kappa'_a}{1 - \kappa'_d/\lambda}  \,.
\end{equation}
Equation (\ref{equationforthepoles}) then reads
\begin{equation}\label{equationforthepoles_zeroed}  
0  = i \biggl(\beta \cos(\beta h) + \frac{\kappa_a}{1 - \kappa_d/\lambda} \sin(\beta h)\biggr),
\end{equation}
which can also be written as
\begin{equation}  \label{eq:system3}
\frac{\tan(\beta h)}{\beta} = - \frac{1 - \kappa_d/\lambda}{\kappa_a} \,.
\end{equation}
We thus get a system of three nonlinear equations
(\ref{eq:system1}, \ref{eq:system2}, \ref{eq:system3}) for the unknown
parameters $\alpha$, $\beta$ and $\lambda$. For each $n$, these equations have infinitely many solutions, and the main practical difficulty in their search is to ensure that they are all found. Doing so analytically is not possible. Thus, while the representation of Eq. (\ref{eq:Jt_3D}) is applicable for the general case, in what follows we focus on two limiting cases where numerical solution is feasible via standard methods.

\subsubsection*{No adsorption on the cylinder wall}

We first discuss the simple limit when there is no adsorption on the cylinder wall, $k'_a = 0$, such that Eq. (\ref{eq:system2}) is reduced to $J_1(\alpha) =
0$, which has infinitely many solutions $\alpha_n = j_{1,n}$, where
$j_{1,n}$ are the zeros of $J_1(z)$, including $\alpha_0 = j_{1,0} = 0$.
When $n = 0$, one has $\lambda = \beta^2$ and Eq. (\ref{eq:system3})
can be written as 
\begin{equation}
\beta \tan(\beta h) = \frac{\kappa_d - \beta^2}{\kappa_a} \,. 
\end{equation}
This equation has infinitely many solutions that we denote $\beta_{0,m}$.
For each $n > 0$, Eq. (\ref{eq:system3}) also has infinitely many
solutions (denoted as $\beta_{n,m}$) but they do not contribute
because the coefficient $c_{n,m}$ is proportional to $J_1(\alpha_n)$
according to Eq. (\ref{eq:cnl}) and thus vanishes.  In other words, when there is no lateral adsorption, the solution $\tilde{J}_{\rm{ab}}(s|r,z)$ and thus $J_{\rm{ab}}(t|r,z)$ do not depend on the radial coordinate, and one retrieves the one-dimensional problem that was solved in ref. \cite{scher2023escape} (see the paragraph below Eq. (9) therein). In this particular case, $\alpha_{n,m} = \alpha_n$ is independent of the second index $m$ and
is actually decoupled from $\beta_{n,m}$.

\subsubsection*{No adsorption on the bottom disk}

Now we discuss another, much more interesting limit when there is no adsorption on the bottom disk.  
If $k_a = 0$, Eq. (\ref{eq:system3}) has infinitely many solutions
$\beta_n h = \pi/2 + \pi n$, with $n = 0,1,2,\ldots$.  We can rewrite
Eq. (\ref{eq:system2}) as
\begin{equation}  \label{eq:auxil1}
\frac{J_0(\alpha)}{\alpha J_1(\alpha)} = \frac{1 - \kappa'_d/(\beta_n^2 + \alpha^2)}{\kappa'_a}  \,.
\end{equation}
The left-hand side decreases piecewise monotonously on the intervals
$(j_{1,l},j_{1,l+1})$, while the right-hand side increases
monotonously from $(1 - \kappa'_d/\beta_n^2)/\kappa'_a$ at $\alpha =
0$ to $1/\kappa'_a$ as $\alpha \to \infty$.  As a consequence, for
each $n$, there are infinitely many solutions that we denote by
$\alpha_{n,l+1}$ for each $l = 0,1,2,\ldots$.

In conventional diffusion problems without desorption, transcendental equations obtained from boundary condition usually admit only real solutions.  In contrast, the desorption mechanism and the consequent $s$-dependence in the Robin boundary condition allow for a purely imaginary solution of Eq. (\ref{eq:auxil1}).  In fact, setting $\alpha = -i \bar{\alpha}$, one gets
\begin{equation}  \label{eq:auxil2}
\frac{I_0(\bar{\alpha})}{\bar{\alpha} I_1(\bar{\alpha})} = \frac{\kappa'_d/(\beta^2 - \bar{\alpha}^2) - 1}{\kappa'_a}  ,
\end{equation}
where $I_\nu(\cdot)$ is the modified Bessel function of the first kind of order $\nu$.
The left-hand side monotonously decreases from $+\infty$ to $0$ as
$\bar{\alpha}$ goes from $0$ to $+\infty$.  In turn, the right-hand
side monotonously increases from $(\kappa'_d/\beta_n^2 - 1)/\kappa'_a$
at $\bar{\alpha} = 0$ to $+\infty$ as $\bar{\alpha}\to \beta_n$, and
then from $-\infty$ to $-1/\kappa'_a$.  As a consequence, there exists
a single solution of this equation on the interval $(0,\beta_n)$, for
each $\beta_n$, that we denote $\bar{\alpha}_n$.  This solution
determines $\alpha_{n,0} = -i \bar{\alpha}_n$ that contributes to the
list of poles.  Note also that this solution results in small
$\lambda_{n,0} = \beta_n^2 - \bar{\alpha}_n^2$; in particular,
$\lambda_{0,0}$ determines the pole with the smallest absolute value,
which in turn determines the asymptotic decay rate of the survival probability \cite{redner2001guide}
\begin{equation}
    S(t|r,z) = 1 - \int_0^t dt J_{\rm{ab}}(t|r,z).
\end{equation}
Since $\beta_n h = \pi/2 + \pi n$, some earlier expressions for the
residues are simplified,
\begin{align}
&\left. \frac{dg_n^{(s)}(H)}{ds} \right|_{s=s_{n,m}} = 
\\
&
\frac{i(-1)^n h L^2}{2D} \biggl(1 + \frac{2\kappa'_a \kappa'_d}{(\kappa'_d - \lambda_{n,m})^2
\bigl(1 + \frac{J_1^2(\alpha_{n,m})}{J_0^2(\alpha_{n,m})}\bigr)}\biggr), \nonumber
\end{align}
such that
\begin{align}
c_{n,m} & = 
\\
&\frac{4D(-1)^n J_1(\alpha_{n,m})}{h L^2 \alpha_{n,m} \Bigl(J_0^2(\alpha_{n,m}) + J_1^2(\alpha_{n,m}) 
+ \frac{2\kappa'_a \kappa'_d J_0^2(\alpha_{n,m})}{(\kappa'_d -\lambda_{n,m})^2}\Bigr)} \,, \nonumber
\end{align}
and the PDF of the escape time becomes
\begin{align} 
& J_{\rm{ab}}(t |r,z)  =   \\  
& \sum\limits_{n,m=0}^\infty e^{-Dt\lambda_{n,m}/L^2} c_{n,m} J_0(\alpha_{n,m} r/L) \beta_{n,m} \cos(\beta_{n,m} z/L) .\nonumber
\end{align}

\subsection{Decay time}\label{sec:holes_decay}

The decay time is determined by the smallest eigenvalue and is hence given by
\begin{equation}\label{decay_tim}
 T  = \frac{L^2}{D\lambda_{0,0}}. 
\end{equation}
\noindent The value of $\lambda_{0,0}$ can be determined numerically,  as described in Sec. \ref{holes_poles}. In turn, the decay time determines the
long-time exponential decay of the survival probability and of the PDF \cite{redner2001guide}:
\begin{equation}
S(t|r,z) \propto e^{-t/T} , \quad J_{\rm ab}(t|r,z) \propto e^{-t/T}  \quad (t\to\infty).
\end{equation}

While one can always find $T$ numerically, in certain cases we can approximate it analytically. For example, let us assume that $L\gg H$ such that $h =H/L \ll 1$. This case corresponds to a very wide and shallow compartment. We recall that for the case of $k_a = 0$ one has $\beta_0 = \pi/(2h)$. A solution
of Eq. (\ref{eq:auxil2}) can be searched (and then validated with simulations) by setting $\bar{\alpha}_0 =
\pi/(2h) - \epsilon$ with $\epsilon \ll 1$.  Substituting this
expression and expanding to the leading order in $\epsilon$, we get
\begin{equation}
\epsilon \approx \frac{\kappa'_d}{\pi/h + 2\kappa'_a \frac{I_0(\pi/(2h))}{I_1(\pi/(2h))}} \,,
\end{equation}
where we note that $\epsilon$ is indeed small when $h \ll 1$, thus the approximation is self-consistent. We then compute $\lambda_{0,0} = \beta_0^2 - \bar{\alpha}_0^2 \approx \pi
\epsilon/h$ and thus
\begin{align} \label{eq:decay} 
T  & \approx \frac{L^2}{D} \, \frac{1 + \frac{2}{\pi} \kappa'_a h \frac{I_0(\pi/(2h))}{I_1(\pi/(2h))}}{\kappa'_d} \\
& = \frac{1}{k'_d} + \frac{2}{\pi} \frac{k'_a H}{k'_d D} \frac{I_0(\pi L/(2H))}{I_1(\pi L/(2H))} \nonumber \\
&\approx \frac{1}{k'_d} \left( 1 + \frac{2}{\pi} \frac{k'_a H}{D}\right)
,\nonumber
\end{align}
where we have used $I_0(\pi L/(2H))/I_1(\pi L/(2H)) \approx 1$ for $h  \ll 1$. Note that in this limit the decay time does not depend on $L$.  


\section{Periodic array of adsorbing nanopillars}\label{sec:pillars}

\begin{figure}[t]
\begin{center}
\includegraphics[width=70mm]{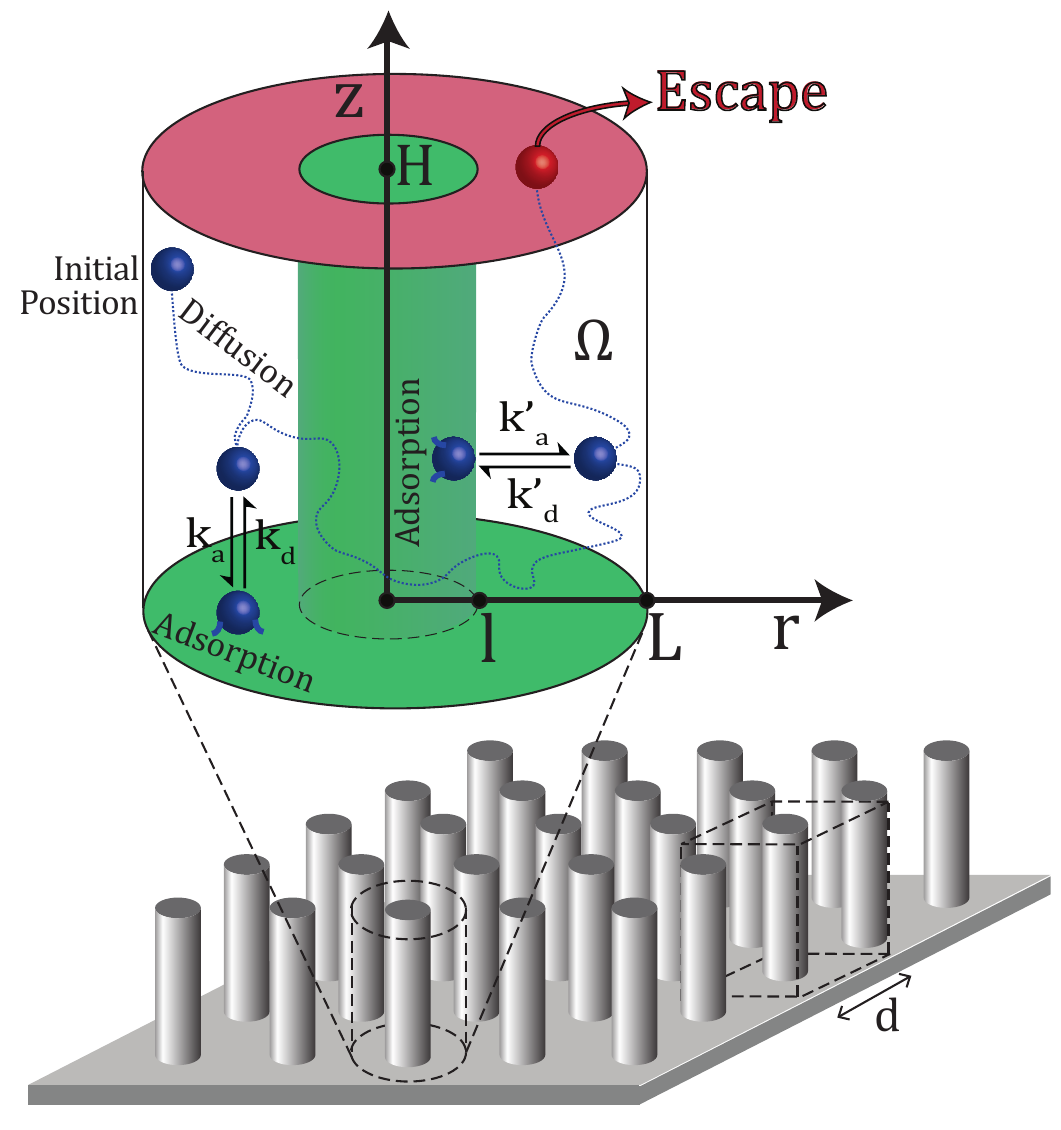}
\end{center}
\caption{
A surface with a periodic array of nanopillars. The periodic cell, drawn in dashed lines, is a rectangular cuboid that we approximate by a cylindrical cell. Such a cylindrical cell is drawn around one of the pillars and enlarged. The radius of the cell is $L=d/\sqrt{\pi}$, where $d$ is the distance between adjacent pillars. The pillar is a cylinder of radius $l$ capped by parallel planes at $z = 0$ and $z =H$.  The top annulus at $z = H$ is absorbing (escape region in red), whereas the bottom annulus at $z = 0$ and the inner cylindrical wall are adsorbing (green). Here, $k_a$ and $k_d$ are the adsorption and desorption constants for the bottom annulus, and $k'_a$ and $k'_d$ are the adsorption and desorption constants for the pillar surface.}
\label{fig:scheme_3D_pillars}
\end{figure}

In this section, we study a different textured surface, which is covered by a periodic array of nanopillars of radius $l$ and height $H$, separated by distance $d$ (Fig. \ref{fig:scheme_3D_pillars}). The survival of a diffusing particle in the presence of absorbing nanopillars was recently studied in \cite{grebenkov2022diffusion,grebenkov2023pillar}. Here, we take a step forward and consider a more challenging situation when the cylindrical walls and the bottom base are {\it adsorbing}. Following the rationales presented in \cite{grebenkov2022diffusion}, we approximate a periodic cell of the structure, a rectangular cuboid, by a cylindrical shell of inner radius $l$ and outer radius $L$, capped by parallel planes at $z = 0$ and $z = H$. In this way, the periodic conditions on the cuboid are replaced by a reflecting boundary condition on the outer cylinder, whose radius $L$ is chosen to be $L=d/\sqrt{\pi}$ to get the same cross-sectional area of the true rectangular cuboid cell, i.e., to preserve the volume of the periodic cell. 

In summary, we consider the escape problem from the above cylindrical shell, in which the top annulus is absorbing, the outer cylinder is reflecting, whereas the inner cylinder and the bottom annulus are adsorbing. The adsorption and desorption rates of the bottom annulus and the inner cylinder can differ. We are interested in finding the PDF of the first-passage time to the top annulus, which can also be thought of as the escape time from the textured surface. We denote this PDF as $J_{\rm{ab}}(t|r,z)$, where $(r,z) \in \Omega$ is the initial location of the particle inside the cylindrical shell. 

Repeating the same considerations as in Eq. (\ref{differential_equation_a})-(\ref{differential_equation_d}) we obtain the boundary value problem
\begin{subequations}
\begin{align} \label{differential_equation_a_2}
(s - D\Delta) \tilde{J}_{\rm{ab}}(s|r,z) & = 0 \quad (r,z \in \Omega), \\ \label{differential_equation_b_2}
\tilde{J}_{\rm{ab}}(s|r,z) & = 1 \quad (z = H) ,\\ \label{differential_equation_c_2}
(-\partial_{z} + q_s) \tilde{J}_{\rm{ab}}(s|r,z) & = 0 \quad (z = 0), \\ \label{differential_equation_d_2}
(-\partial_r + q'_s) \tilde{J}_{\rm{ab}}(s|r,z) & = 0 \quad (r = l),  \\ \label{differential_equation_e_2}
\partial_r \tilde{J}_{\rm{ab}}(s|r,z) & = 0 \quad (r = L),
\end{align}
\end{subequations}
where $\Delta = \partial_{r}^2 + (1/r)\partial_{r} + \partial_{z}^2$ is the
Laplace operator in cylindrical coordinates (without the angular part). The surfaces are characterized by the parameters $q_s$ and $q'_s$ that were defined in Eq. (\ref{eq:qs}),
where $k_a$ and $k_d$ are the adsorption and desorption constants for the
bottom annulus, and $k'_a$ and $k'_d$ are the adsorption and desoprtion
constants for the inner cylindrical surface.

\subsection{Solution in Laplace domain}\label{pillars_Laplace}

In analogy to Sec. \ref{sec:holes}, we search the solution for Eq. (\ref{differential_equation_a_2})  under the boundary conditions (\ref{differential_equation_b_2})-(\ref{differential_equation_e_2}) as 
\begin{equation}  \label{eq:tildeJ_general_2}
\tilde{J}_{\rm{ab}}(s|r,z) =  \sum\limits_{n=0}^\infty c_n^{(s)} \omega_0(\alpha_n^{(s)},\bar{r}) \frac{g_n^{(s)}(z)}{g_n^{(s)}(H)} \,,
\end{equation}
where $\bar{r}=r/L$, with $g_n^{(s)}(z)$ and $\hat{\alpha}_n^{(s)}$ being defined in Eqs. (\ref{eq:g_def}, \ref{eq:hat_alpha}). The prefactors in $g_n^{(s)}(z)$ 
were determined to ensure the boundary conditions (\ref{differential_equation_b_2})-(\ref{differential_equation_c_2}).
We further introduce
\begin{equation}   \label{omega_def}
\omega_\nu(\alpha_n^{(s)},\bar{r}) \equiv
J_1(\alpha_n^{(s)}) Y_\nu(\alpha_n^{(s)} \bar{r}) - Y_1(\alpha_n^{(s)}) J_\nu(\alpha_n^{(s)}\bar{r}),
\end{equation}
where $Y_\nu(\cdot)$ is the Bessel function of the second kind of order $\nu$, such that the condition (\ref{differential_equation_e_2}) is satisfied. Indeed, the relation $\partial_r \omega_0(\alpha_n^{(s)},\bar{r}) = - \alpha_n^{(s)} \omega_1(\alpha_n^{(s)},\bar{r})/L$ implies $\partial_r \omega_0(\alpha_n^{(s)},\bar{r})|_{\bar{r}=1}=0$. 
From the boundary condition (\ref{differential_equation_d_2}) we find that
$\alpha_n^{(s)}$ satisfy the transcendental equation
\begin{equation}  \label{eq:alphan_eq_2}
\alpha_n^{(s)} \frac{\omega_1(\alpha_n^{(s)}, \rho)}{\omega_0(\alpha_n^{(s)},\rho)} = -q'_s L,
\end{equation}
where $\rho = l/L$. For any $s\geq 0$, there are infinitely many solutions that we enumerate by $n = 0,1,2,\ldots$ in an increasing order.
The unknown
coefficients $c_n^{(s)}$ are found by multiplying the boundary condition (\ref{differential_equation_b_2}) by $r\omega_0(\alpha_n^{(s)},r)$ and
integrating over $r$ from $l$ to $L$. This gives
\begin{align} \label{to_impose_b_2}
&\frac{c_k^{(s)}}{2} \times \\
& \left\{ \left[ \omega_0(\alpha_n^{(s)},1)\right]^2 - 
\left[\rho\omega_0(\alpha_n^{(s)},\rho)\right]^2
 - 
\left[\rho\omega_1(\alpha_n^{(s)},\rho)\right]^2
 \right\}
 \nonumber \\
 &= -\frac{\rho}{\alpha_k^{(s)}} \omega_1(\alpha_n^{(s)},\rho), \nonumber
\end{align}
from which we get
\begin{align}\label{c_coef_2}
&c_n^{(s)} = \\
&\frac{2\rho \omega_1(\alpha_n^{(s)},\rho)/\alpha_n^{(s)}  }{   
\left[\rho\omega_1(\alpha_n^{(s)},\rho) \right]^2 + \left[\rho\omega_0(\alpha_n^{(s)},\rho)\right]^2 - \left[ \omega_0(\alpha_n^{(s)},1)\right]^2} \,. \nonumber
\end{align}
To obtain the left-hand side of Eq. (\ref{to_impose_b_2}) we used the orthogonality of the Bessel functions together with
\begin{align}
&c_k^{(s)}\int_\rho^1 d \bar{r} \, \bar{r} \, \left[\omega_0(\alpha_n^{(s)},\bar{r})\right]^2  =
\\
&\frac{c_k^{(s)}}{2} \left\{  \left[\bar{r}\omega_0(\alpha_n^{(s)},\bar{r})\right]^2 + \left[r\omega_1(\alpha_n^{(s)},\bar{r})\right]^2   \right\}_\rho^1 \, ,
 \nonumber 
\end{align}
and noted that $\omega_1(\alpha_n^{(s)},1)=0$.
To obtain the right-hand side of Eq. (\ref{to_impose_b_2}), we used
\begin{equation}
\int\limits_\rho^1 d\bar{r} \, \bar{r} \,\omega_0(\alpha_n^{(s)},\bar{r}) =  -\frac{\rho}{\alpha_k^{(s)}} \omega_1(\alpha_n^{(s)},\rho).
\end{equation}

\noindent To facilitate further analysis, we use the dimensionless quantities in Table I. 


\subsection{Mean Escape Time}\label{pillar_mean}

In this section, we compute the mean escape time by analyzing the asymptotic
behavior of $\tilde{J}_{\rm{ab}}(s|r,z)$ as $s\to 0$. We employ the same procedure that was described in Sec. \ref{holes_mean}. 

In the spectral expansion
(\ref{eq:tildeJ_general_2}), we first analyze the term $n = 0$ and then
discuss the other terms with $n > 0$. As $s \to 0$, one has $q'_s \to 0$  (with an asymptotic form $\frac{\kappa'_a}{\kappa'_d}\frac{L}{D}s$) and so, according to Eq. (\ref{eq:alphan_eq_2}), $\alpha_0^{(s)} \to 0$.  Using
the Taylor expansion of the Bessel functions in
Eq. (\ref{omega_def}), one gets
\begin{align}
&\omega_0(\alpha_n^{(s)},\bar{r}) = \\
& \frac{2}{\pi \alpha_n^{(s)}} + \alpha_n^{(s)} \frac{1 - \bar{r}^2 + 2 \ln(\bar{r}) }{2\pi } +  O((\alpha_n^{(s)})^2)\,, \nonumber
\end{align}
and
\begin{equation}
\omega_1(\alpha_n^{(s)},\bar{r}) \approx -\frac{1 - \bar{r}^2}{\pi \bar{r} } +  O((\alpha_n^{(s)})^2).
\end{equation}
Therefore, for $s \to 0$ we obtain
\begin{equation}
\alpha_0^{(s)} \approx \sqrt{s} \sqrt{2\kappa'_a/k'_d}\sqrt{\frac{\rho}{1-\rho^2}} \quad (s\to 0)\,.
\end{equation}
As a consequence, we have
\begin{align}
& c_0^{(s)} \approx \\
& \frac{ \pi}{2}\alpha_0^{(s)} \left( 1 + \frac{\rho}{1-\rho^2}\frac{\kappa'_a}{k'_d} \frac{1 - \rho^4 + 4\rho^2 \ln(\rho)}{4 (1-\rho^2)}  s \right) +       O(s^{5/2}). \nonumber
\end{align}
We also get $\hat{\alpha}_0^{(s)} \approx  \sqrt{s} \sqrt{L^2/D}\sqrt{1 +
2(\rho/(1-\rho^2))\kappa'_a/\kappa'_d}$ in the leading order, from which we obtain
\begin{align}
&\frac{g_0^{(s)}(z)}{g_0^{(s)}(H)} = 1 - \\
 & \hspace{3ex}  \biggl(\frac{\kappa_a}{k_d} \frac{H-z}{L} + \biggl[\frac{L^2}{2D} + \frac{\rho}{1-\rho^2}\frac{\kappa'_a}{k'_d}\biggr]
\frac{H^2-z^2}{L^2}\biggr) s + O(s^2). \nonumber
\end{align}

Let us now consider the terms with $n > 0$.  Denoting the left-hand
side of Eq. (\ref{eq:alphan_eq_2}) as 
\begin{equation}\label{70}
F_1(x) = x \frac{\omega_1(x,\rho)}{\omega_0(x,\rho)} ,
\end{equation}
we Taylor expand
\begin{equation} \label{F_taylor_2}
F_1(\alpha_n^{(s)}) \approx F_1(\alpha_n^{(0)}) + F_1'(\alpha_n^{(0)}) (\alpha_n^{(s)} - \alpha_n^{(0)}) .
\end{equation}
According to Eq. (\ref{eq:alphan_eq_2}), $F_1(\alpha_n^{(s)}) = - q'_s L \to
0$ as $s \to 0$ and thus $\alpha_n^{(0)}$ are the zeros of the function $\omega_1(x,\rho)$ defined in Eq. (\ref{omega_def}). Taking the derivative of Eq. (\ref{70}) and plugging in $x = \alpha_n^{(0)}$ we find 
\begin{align}
 F_1'(\alpha_n^{(0)}) =
 \alpha_n^{(0)}\rho - \frac{4}{\pi^2 \alpha_n^{(0)} \rho \left[\omega_0(\alpha_n^{(0)},\rho)\right]^2} . 
\end{align}

\begin{figure}
\begin{center}
\includegraphics[width=70mm]{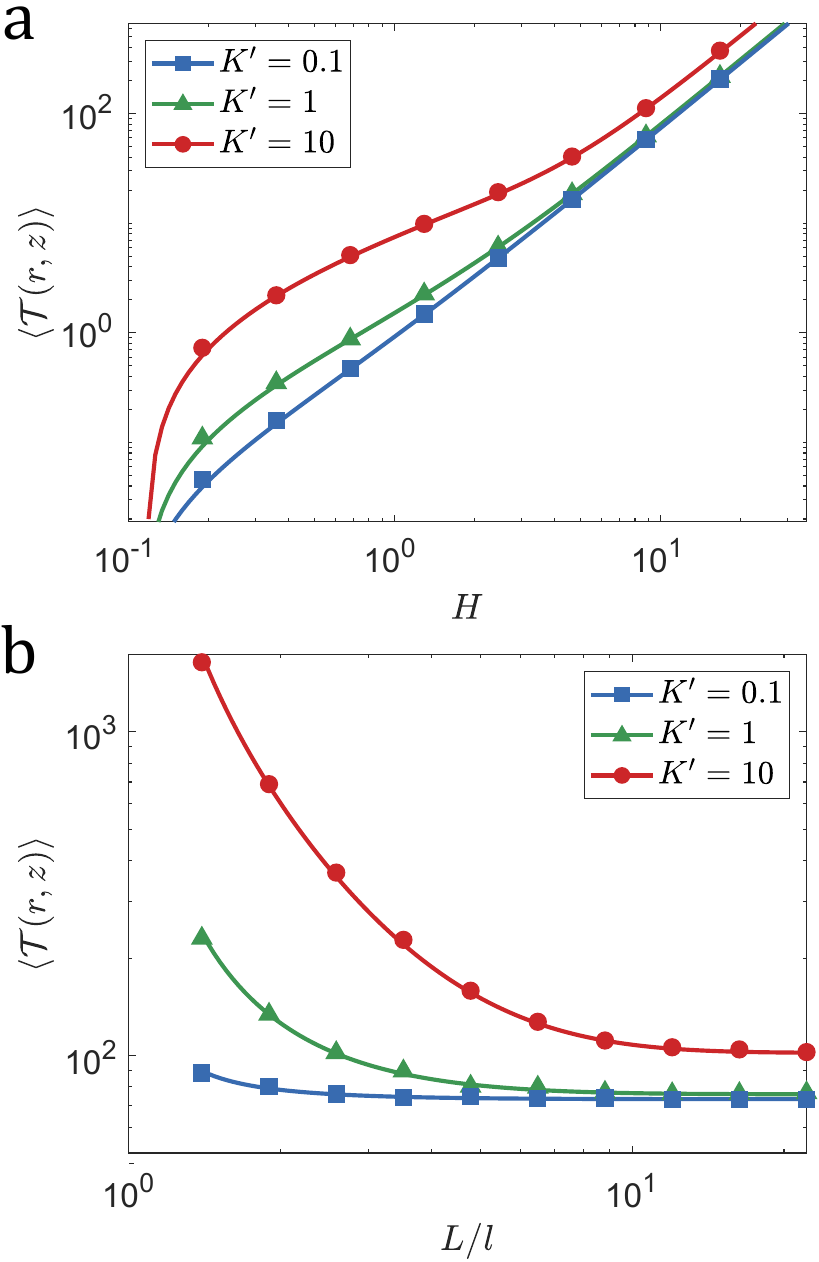}
\end{center}
\caption{
Mean escape time from a surface textured by an array of pillars (Fig. \ref{fig:scheme_3D_pillars}).
Solid lines are drawn using the exact solution of Eq. (\ref{eq:Tmean_2}). Each line represents a different adsorption equilibrium constant for the pillar, $K'=k'_a/k'_d$, where we set $k'_d=1$ and vary $k'_a$ accordingly. Marker symbols represent the mean escape time of $10^4$ particles simulated according to the protocol in Appendix D of Ref. \cite{scher2023escape}, with simulation time step $\Delta t=10^{-4}$. We set $D = 0.7$, $l=0.9$, $k_a =
0.7$, $k_d =
6.1$, $z=0.1$ and $r=1$. The mean escape time is plotted as function of 
(a) the height of the pillars $H$, where we set $L=5$; (b) the radius of the unit cell $L$ (divided by the radius of the pillars $l$), where we set $H =10$.}\label{fig:mean_pillars}
\end{figure}

\noindent Comparing the right-hand side of Eqs. (\ref{F_taylor_2}) and  (\ref{eq:alphan_eq_2}) in the limit $s \to 0$, we obtain 
\begin{equation} \label{alpha_n_taylor_2}
\alpha_n^{(s)} = \alpha_n^{(0)} - \frac{\kappa'_a}{k'_d }\frac{1}{F_1'(\alpha_n^{(0)})} s + O(s^2).
\end{equation}
Similarly, setting
\begin{align}\label{74}
 &F_2(x) =
 \frac{2\rho \omega_1\left(x,\rho\right)/x }{   
\left[\rho\omega_1\left(x,\rho\right)\right]^2 + \left[\rho\omega_0\left(x,\rho\right)\right]^2 - \left[ \omega_0(x,1)\right]^2} , 
\end{align}
we have $c_n^{(s)}  = F_2(\alpha_n^{(s)})$ according to  Eq. (\ref{c_coef_2}). Taking the derivative of Eq. (\ref{74}) and plugging in $x = \alpha_n^{(0)}$ we find 
\begin{widetext}
\begin{align}
&F_2'(\alpha_n^{(0)}) =  \frac{J_0(\alpha_n^{(0)})Y_{1}(\alpha_n^{(0)}\rho) + Y_{2}(\alpha_n^{(0)})J_1(\alpha_n^{(0)}\rho) -\rho\left[ Y_{1}(\alpha_n^{(0)})J_0(\alpha_n^{(0)}\rho)  +J_{1}(\alpha_n^{(0)})Y_2(\alpha_n^{(0)}\rho)  \right] }{  -\frac{2}{\pi^2 \alpha_n^{(0)} \rho}  + \frac{\alpha_n^{(0)} \rho}{2} \left[ \omega_0(\alpha_n^{(0)},\rho)\right]^2}  
\end{align}
\end{widetext}
and using Eq. (\ref{alpha_n_taylor_2}) we get
\begin{align} 
c_n^{(s)}  &=  F_2(\alpha_n^{(0)}) + F'_2(\alpha_n^{(0)}) (\alpha_n^{(s)} - \alpha_n^{(0)}) + O(s^2) \\
& = -\frac{F'_2(\alpha_n^{(0)})}{F_1'(\alpha_n^{(0)})} \frac{\kappa'_a}{k'_d  } s  + O(s^2), \nonumber
\end{align}
and
\begin{equation}
\hat{\alpha}_n^{(s)} = \alpha_n^{(0)} + \left( \frac{1}{2 \alpha_n^{(0)}} - \frac{\kappa'_a}{\kappa'_d  F_1'(\alpha_n^{(0)})} \right) \frac{L^2 }{D} s + O(s^2).
\end{equation}
As a consequence, we find that to the leading order in $s$
\begin{equation}
\frac{g_n^{(s)}(z)}{g_n^{(s)}(H)}  \approx \frac{\cosh(\alpha_n^{(0)} \bar{z})}{\cosh(\alpha_n^{(0)} h)},
\end{equation}
and
\begin{equation}
\omega_0(\alpha_n^{(s)},\bar{r}) \approx \omega_0(\alpha_n^{(0)},\bar{r}),
\end{equation}
with $\bar{z} = z/L$.
Substituting these expressions into Eq. (\ref{eq:tildeJ_general_2}), we
get
\begin{equation}
\tilde{J}_{\rm{ab}}(s|r,z) = 1 - s\langle \mathcal{T} (r,z) \rangle + O(s^2),
\end{equation}
where
\begin{widetext}
\begin{align}  \label{eq:Tmean_2}
\langle \mathcal{T} (r,z) \rangle  &= \frac{H^2-z^2}{2D} +\frac{k_a(H-z)}{k_d D}  \\  \nonumber
&  +\frac{k'_a}{k'_d}\frac{L}{D}\left[ \frac{\rho}{1 - \rho^2} \left(h^2 - \bar{z}^2 -  \frac{ 1 - \bar{r}^2 + 2\ln(\bar{r})}{2} - \frac{ 1 - \rho^4 + 4\rho^2 \ln(\rho)}{4 (1-\rho^2)} 
\right) +  \sum\limits_{n=1}^\infty \omega_0(\alpha_n^{(0)},\bar{r})  
\frac{\cosh(\alpha_n^{(0)} \bar{z})}{\cosh(\alpha_n^{(0)} h)} \frac{F'_2(\alpha_n^{(0)})}{F_1'(\alpha_n^{(0)})}  \right]  
\end{align}
\end{widetext}
is the mean escape time, where $h=H/L$ is the dimensionless cylinder's height. In Fig. \ref{fig:mean_pillars}, we plot the mean escape time from a surface textured by an array of pillars as given by Eq. (\ref{eq:Tmean_2}), with varying pillar height and varying inter-pillar distance.

As in Sec. \ref{holes_mean}, in the limit $k'_a \to 0$ or $k'_d \to \infty$, when the pillars are not sticky but inert, we retrieve the mean escape time for a one-dimensional box with a sticky surface \cite{scher2023escape} 
\begin{equation}
\langle \mathcal{T} (r,z) \rangle =
\frac{1}{2D}(H^2-z^2) + \frac{k_a}{k_d D}(H-z),
\end{equation}
which is identical to Eq. (\ref{24}).

The average of Eq. (\ref{eq:Tmean_2}) over the cross-section at height $z$ yields 
\begin{align}\label{eq:Tmean_average_2}
\langle\overline{\mathcal{T}}(z)\rangle & = \frac{2\pi}{\pi (L^2-l^2)} \int\limits_l^L dr \, r\, \langle \mathcal{T} (r,z) \rangle \\    
& =    \frac{k_a}{k_d}  \frac{H-z}{D} +\left(1+ \frac{2 k'_a}{k'_d L} \frac{\rho}{1 - \rho^2} \right) \frac{H^2-z^2}{2D}  \,,  \nonumber
\end{align}
where we used $\int_l^L dr \, \omega_0(\alpha_n^{(0)},\bar{r}) = 0$. Further averaging over $z$, we obtain 
\begin{align}\label{mean_uniform_2}  
\langle \mathcal{T}_{u}\rangle  = \frac{H^2}{3D} \left( 1+\frac{k_a}{k_d}  \frac{3}{2H} + \frac{k'_a}{k'_d} \frac{\rho}{1-\rho^2} \frac{2}{L} \right),
\end{align}
where the subscript `$u$' denotes a uniform distribution of the initial position.

\subsection{Solution in time domain}\label{pillars_time_domain}

The solution in time domain can be found via the residue theorem.  The computation is very similar to the case of the capped cylinder in Sec. \ref{holes_time_domain}; here, we
differentiate Eq. (\ref{eq:alphan_eq_2}) to get
\begin{align}  \label{eq:alp_dalp_pillar}
&\alpha_n^{(s)} \frac{d\alpha_n^{(s)}}{ds} = -\frac{\kappa'_a k'_d}{(k'_d + s)^2} \times  \Bigg( \frac{\omega_1(\alpha_n^{(s)}, \rho)}{\alpha_n^{(s)}\omega_0(\alpha_n^{(s)},\rho)} \\ 
&  + \frac{\omega_0(\alpha_n^{(s)},\rho)\frac{d\omega_1(\alpha_n^{(s)},\rho)}{d\alpha_n^{(s)}} - \omega_1(\alpha_n^{(s)},\rho)\frac{d\omega_0(\alpha_n^{(s)},\rho)}{d\alpha_n^{(s)}}}{\left[\omega_0(\alpha_n^{(s)},\rho)\right]^2}\Bigg)^{-1} .\nonumber
\end{align}
Overall, we obtain
\begin{align} \label{eq:Jt_3D_exact_2}
& J_{\rm{ab}}(t|x,z)  = \sum\limits_{n,m=0}^\infty c_{n,m} e^{-Dt\lambda_{n,m}/L^2} \omega_0(\alpha_{n,m} r/L) \\  
& \times \biggl(\beta_{n,m} \cos(\beta_{n,m} z/L) - \frac{\kappa_a \lambda_{n,m}}{\kappa_d - \lambda_{n,m}} \sin(\beta_{n,m} z/L)\biggr), \nonumber
\end{align}
where $\alpha_{n,m} = \alpha_n^{(s_{n,m})}$, $\lambda_{n,m} =
\alpha_{n,m}^2 + \beta_{n,m}^2 = - s_{n,m} L^2/D$,
\begin{align} 
c_{n,m} & = \frac{i c_n^{(s_{n,m})}}{\frac{dg_n^{(s)}(H)}{ds}\bigr|_{s=s_{n,m}}} = \frac{i}{\frac{dg_n^{(s)}(H)}{ds} \bigr|_{s=s_{n,m}}} \times \\
& \frac{2\rho\omega_1\left(\alpha_{n,m},\rho\right)/\alpha_{n,m}  }{  
\left[\rho\omega_1\left(\alpha_{n,m},\rho\right) \right]^2 + \left[\rho\omega_0\left(\alpha_{n,m},\rho\right)\right]^2 - \left[ \omega_0(\alpha_{n,m},1)\right]^2}  , \nonumber
\end{align}
and $\frac{dg_n(H)}{ds}|_{s=s_{n,m}}$ is given by
Eq. (\ref{eq:dgnH_ds}), in which $\alpha_n^{(s)}
\frac{d\alpha_n^{(s)}}{ds}$ is substituted from
Eq. (\ref{eq:alp_dalp_pillar}).



\subsection{Poles}\label{pillars_poles}

The poles of $\tilde{J}_{\rm{ab}}(s|x,z)$ are determined by zeros of
$g_n^{(s)}(H)$, as previously (see Sec. \ref{holes_poles}).  We use the former notations:
$\alpha_n^{(s)} = \alpha$, $\hat{\alpha}_n^{(s)} = i\beta$, and
$\alpha^2 + \beta^2 = \lambda = - s L^2/D$.  In this case,
Eq. (\ref{eq:alphan_eq_2}) reads
\begin{equation}  \label{eq:system1_2D_2}
 \frac{\omega_0(\alpha, \rho)}{\alpha \omega_1(\alpha, \rho)} = - \frac{1 - \kappa'_d/(\alpha^2 + \beta^2)}{\kappa'_a} .
\end{equation}
We focus on the case $k_a = 0$, for which $\beta_n h = \pi/2 + \pi n$. For any fixed $\beta_n$, the left-hand side of Eq. (\ref{eq:system1_2D_2}) increases piecewise monotonously from $-\infty$ to $+\infty$ on the intervals $(\alpha_n^{(0)}, \alpha_{n+1}^{(0)})$, while the right-hand side is a monotonously decreasing function of $\alpha$.  As a consequence, there is a single solution on each interval $(\alpha_n^{(0)}, \alpha_{n+1}^{(0)})$ that we denote as $\alpha_{n,m+1}$, for $m = 0,1,2,\ldots$.  In addition, there is a purely imaginary solution, which can be found by setting 
$\alpha = -i\bar{\alpha}$, with $\bar{\alpha}$ satisfying
\begin{align}  \label{eq:auxil3_2}
&\frac{1}{\bar{\alpha}}\frac{I_1(\bar{\alpha}) K_0(\bar{\alpha} \rho) + K_1(\bar{\alpha}) I_0(\bar{\alpha} \rho)}{K_1(\bar{\alpha}) I_1(\bar{\alpha} \rho) - I_1(\bar{\alpha}) K_1(\bar{\alpha} \rho) }  \nonumber\\
& = \frac{1 - \kappa'_d/(\beta^2 - \bar{\alpha}^2) }{\kappa'_a}  ,
\end{align}
where $I_\nu(\cdot)$ and $K_\nu(\cdot)$ are the modified Bessel functions of the first and second kind of order $\nu$.
The left-hand side monotonously increases from $-\infty$ to $0$ as
$\bar{\alpha}$ goes from $0$ to $+\infty$,
whereas the right-hand side for any fixed $\beta_n$ decreases
monotonously on $(0,\beta_n)$ from $(\kappa'_d/\beta_n^2 -
1)/\kappa'_a$ to $-\infty$, and on $(\beta_n,+\infty)$ from $\infty$
to $1/\kappa'_a$.  As a consequence, there exists only one solution
on the interval $(0,\beta_n)$ that we denote $\bar{\alpha}_n$.  This
solution determines $\alpha_{n,0} = - i\bar{\alpha}_n$ that
contributes to the list of poles.

\subsection{Decay time}\label{sec:pillars_decay}

The general discussion in Sec. \ref{sec:holes_decay} is valid here. Let us find the approximation for the decay time $T$ in the the limit $k_a=0$ and $h =H/L \ll 1$ such that $\beta_0 = \pi/(2h) \gg 1$. A solution of Eq. (\ref{eq:auxil3_2}) can be searched (and then validated with simulations) by setting $\bar{\alpha}_0 = \pi/(2h) - \epsilon$ with $\epsilon \ll 1$.  Substituting this expression and expanding to the leading order in $\epsilon$, we get
\begin{equation}
 \epsilon \approx \left[\frac{\pi}{h \kappa'_d}+\frac{2 \kappa'_a}{\kappa'_d}\frac{I_1(\frac{\pi }{2 h}) K_0(\frac{\pi \rho}{2 h}) + K_1(\frac{\pi }{2 h}) I_0(\frac{\pi \rho}{2 h}) }{ I_1(\frac{\pi }{2 h}) K_1(\frac{\pi \rho}{2 h}) - K_1(\frac{\pi }{2 h}) I_1(\frac{\pi \rho}{2 h})} \right]^{-1}  ,
\end{equation}
where we note that $\epsilon$ is indeed small when $h \ll 1$, thus the approximation is self-consistent. We then compute $\lambda_{0,0} = \beta_0^2 - \bar{\alpha}_0^2 \approx \pi
\epsilon/h$ and find 
\begin{align} \label{eq:approx_2}
T & = \frac{L^2}{D\lambda_{0,0}}  \nonumber
\\  
& \approx \frac{1}{k'_d} + \frac{2 h \kappa'_a}{ \pi k'_d }\frac{I_1(\frac{\pi }{2 h}) K_0(\frac{\pi \rho}{2 h}) + K_1(\frac{\pi }{2 h}) I_0(\frac{\pi \rho}{2 h})}{ I_1(\frac{\pi }{2 h}) K_1(\frac{\pi \rho}{2 h}) - K_1(\frac{\pi }{2 h}) I_1(\frac{\pi \rho}{2 h})}  .
\end{align}
To proceed, we note that when $h \ll 1$, one has $ I_\nu(\frac{\pi}{2h}) \approx
\sqrt{h} e^{\pi/(2h)}/\pi$ whereas $ K_\nu(\frac{\pi}{2h}) \approx \sqrt{h}
e^{-\pi/(2h)}$ and is thus negligible. We then get 
\begin{align} \label{eq:decay_2}
   T &\approx \frac{1}{k'_d} + \frac{2 H k'_a}{\pi D k'_d} \, \frac{K_0(\pi l/(2H))}{K_1(\pi l/(2 H))}    \\
     &\approx  \frac{1}{k'_d} \left( 1 + \frac{2}{\pi} \frac{k'_a H}{D} \right), \nonumber
\end{align}
where we noted that $\pi \rho/(2 h) = \pi l/(2H)$ and used $K_0(\pi l/(2H))/K_1(\pi l/(2 H)) \approx 1$ for $h  \ll 1$. We thus see that in this limit the decay time identifies with the decay time in Eq. (\ref{eq:decay}), and that it does not depend on $L$. 


\section{Adsorbing grooved surface}
\label{sec:grooves}

\begin{figure}
\begin{center}
\includegraphics[width=70mm]{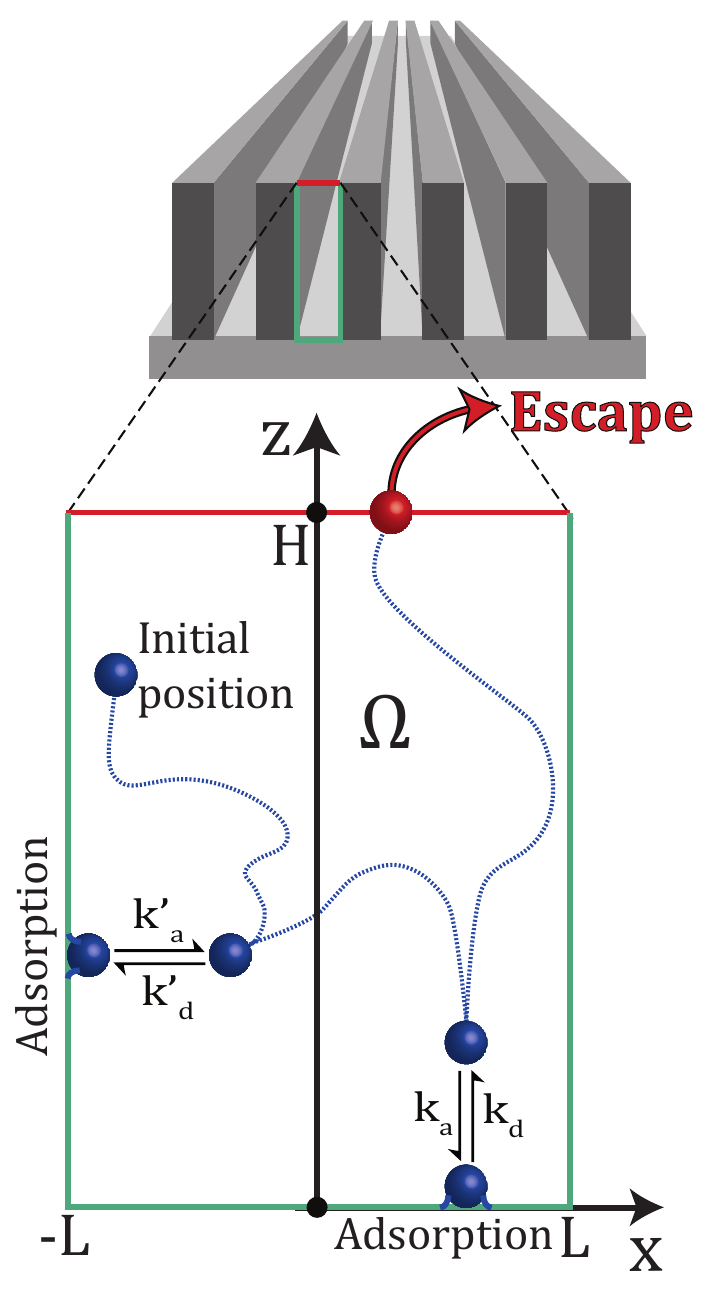}
\end{center}
\caption{
Adsorbing grooved surface. The walls separating the grooves are of height $H$ and the distance between any two walls is $2L$. One of the grooves is enlarged, and the problem is effectively the  escape from a two-dimensional rectangular compartment. The top side at $z = H$ is absorbing (an escape region in red), whereas
the other three sides are adsorbing (green). Here, $k_a$ and $k_d$ are the adsorption and desorption constants for the bottom edge, and $k'_a$ and $k'_d$ are the adsorption and desorption constants for the left and right edges.}
\label{fig:scheme_2D_grooves}
\end{figure}

\begin{figure}
\begin{center}
\includegraphics[width=85mm]{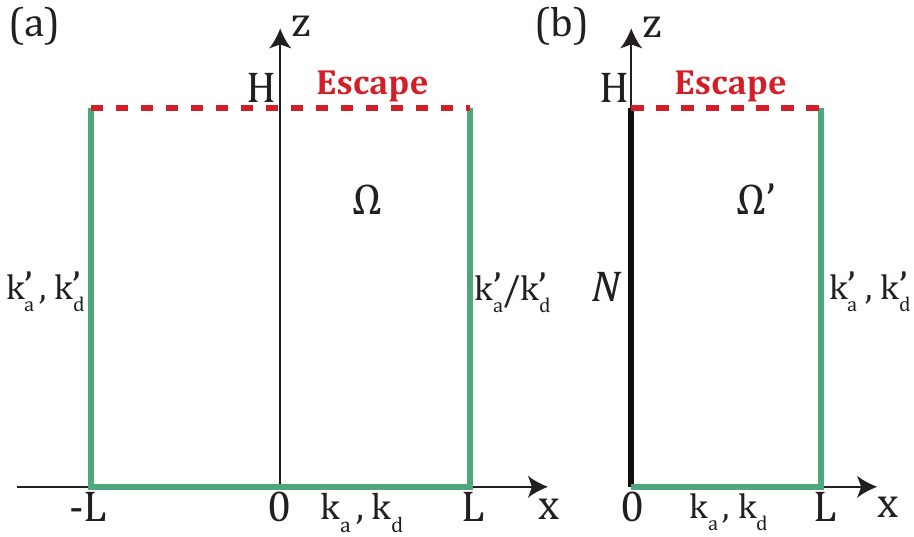}
\end{center}
\caption{
{\bf(a)} A schematic illustration of a rectangular domain $\Omega =
(-L,L)\times (0,H) \subset \R^2$ with one absorbing edge (an escape region on the top), and three adsorbing edges with reversible binding kinetics characterized by $k_a$ and $k_d$ (bottom) and $k'_a$ and $k'_d$ (left and right). {\bf (b)} An equivalent twice smaller domain $\Omega'=(0,L)\times (0,H)$ with a reflecting edge replacing the adsorbing edge on the left (a Neumann boundary condition is denoted by $N$).}
\label{fig:scheme_2D}
\end{figure}

We consider a grooved surface, as illustrated in Fig. \ref{fig:scheme_2D_grooves}.  This problem is equivalent to diffusion with a diffusion coefficient $D$ in a
rectangular domain $\Omega=(-L,L)\times (0,H)$. The top
edge of the domain is absorbing (with Dirichlet boundary condition), and the three other edges are {\it adsorbing}, with reversible binding. As in the previous examples, we allow for different adsorption kinetics on the bottom edge.  We search the probability density function $J_{\rm{ab}}(t|x,z)$ of the escape time through the top edge. Note that the problem of escape from the domain $\Omega$ is equivalent to an escape from a twice smaller domain $\Omega' = (0,L)\times (0,H)$ where the left edge is reflecting (with Neumann boundary condition), see Fig. \ref{fig:scheme_2D}. We thus focus on the latter setting.

Repeating the same considerations as in Eq. (\ref{differential_equation_a})-(\ref{differential_equation_d}) we obtain the boundary value problem
\begin{subequations}
\begin{align} \label{differential_equation_a_3}
(s - D\Delta) \tilde{J}_{\rm{ab}}(s|x,z) & = 0 \quad (x,z \in \Omega'), \\ \label{differential_equation_b_3}
\tilde{J}_{\rm{ab}}(s|x,z) & = 1 \quad (z = H) ,\\ \label{differential_equation_c_3}
(-\partial_{z} + q_s) \tilde{J}_{\rm{ab}}(s|x,z) & = 0 \quad (z = 0), \\ \label{differential_equation_d_3}
-\partial_x  \tilde{J}_{\rm{ab}}(s|x,z) & = 0 \quad (x = 0),  \\ \label{differential_equation_e_3}
(\partial_x + q'_s) \tilde{J}_{\rm{ab}}(s|x,z) & = 0 \quad (x = L),
\end{align}
\end{subequations}
where $\Delta = \partial_{x}^2 + \partial_{z}^2$ is the
Laplace operator in Cartesian coordinates. The surfaces are characterized by the parameters $q_s$ and $q'_s$ that were defined in Eq. (\ref{eq:qs}), where $k_a$ and $k_d$ are the adsorption and desorption constants for the bottom edge, and $k'_a$ and $k'_d$ are the adsorption and desoprtion constants for the right edge. 

\subsection{Solution in Laplace domain}\label{grooves_Laplace}

The solution of Eq. (\ref{differential_equation_a_3}) under the boundary conditions (\ref{differential_equation_b_3})-(\ref{differential_equation_e_3}) is 
\begin{equation}  \label{eq:Jtilde_eq_2D}
\tilde{J}_{\rm{ab}}(s|x,z) = \sum\limits_{n=0}^\infty c_n^{(s)} \, \cos(\alpha_n^{(s)} x/L) \frac{g_n^{(s)}(z)}{g_n^{(s)}(H)} \,,
\end{equation}
where $g_n^{(s)}(z)$ and $\hat{\alpha}_n^{(s)}$ were defined in Eqs. (\ref{eq:g_def}, \ref{eq:hat_alpha}). The prefactors in $g_n^{(s)}(z)$ 
were determined by the boundary conditions (\ref{differential_equation_b_3})-(\ref{differential_equation_c_3}).

We have used the reflecting boundary condition (\ref{differential_equation_d_3}) to determine the form of the $x$-dependent part in Eq. (\ref{eq:Jtilde_eq_2D}). 
From the boundary condition (\ref{differential_equation_e_3}) we find that
$\alpha_n^{(s)}$ satisfy the transcendental equation
\begin{equation}  \label{eq:alpha_n_eq_2D}
\alpha_n^{(s)} \tan(\alpha_n^{(s)}) = q'_s L.
\end{equation}
For any $s\geq 0$, there are infinitely many solutions that we enumerate by $n = 0,1,2,\ldots$ in an increasing order.
The unknown
coefficients $c_n^{(s)}$ are found by multiplying the boundary condition (\ref{differential_equation_b_3}) by $\cos(\alpha_k^{(s)} x/L)$ and
integrating over $x$ from $0$ to $L$. This gives
\begin{equation} 
c_n^{(s)}  = \frac{2\sin(\alpha_n^{(s)})}{\alpha_n^{(s)} \Bigr(1 + \frac{\sin(2\alpha_n^{(s)})}{2\alpha_n^{(s)}}\Bigl)}.
\end{equation}

To facilitate further analysis, we use the dimensionless quantities in Table I. In the solution process, we will employ the same procedure that was described in Sec. \ref{sec:holes}. We will thus skip some of the details of the calculation.

\begin{figure*}[t]
\begin{center}
\includegraphics[width=\linewidth]{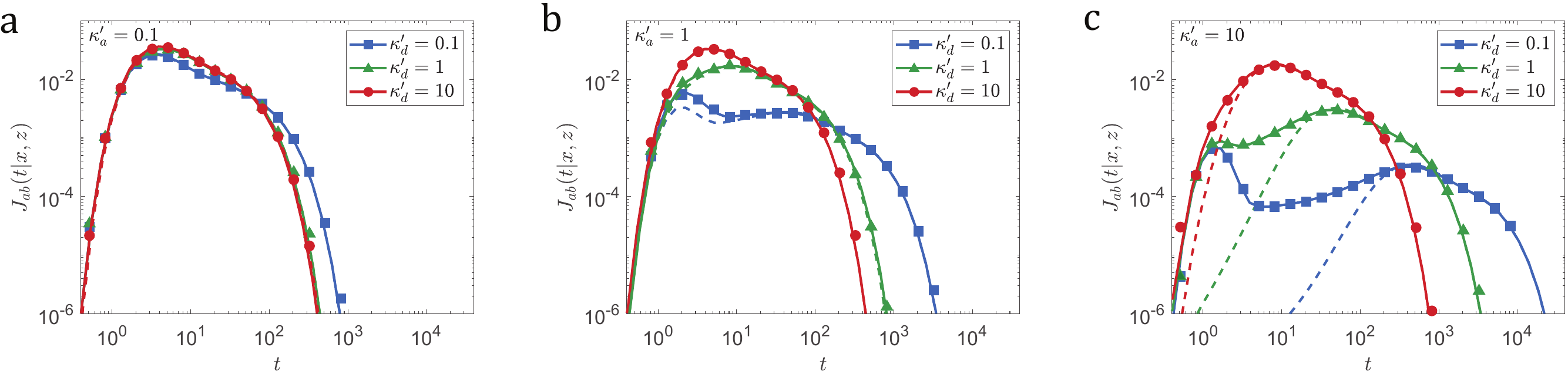}
\end{center}
\caption{PDF $J_{\rm{ab}}(t|x,z)$ of the escape time from a groove with $D = 1$, $L = 1$, $H =10$, $k_a = 0$, for three different values of $\kappa'_d$ (see the legend), and $\kappa'_a = 0.1$ (panel a), $\kappa'_a = 1$ (panel b), and $\kappa'_a = 10$ (panel c).  Solid lines give the exact solution from Eq. (\ref{eq:Jt_2D_exact}), while dashed lines represent the two-state switching diffusion approximation. Marker symbols give estimates based on $10^6$ particles whose motion was simulated according to the protocol in Appendix D of Ref. \cite{scher2023escape}, with simulation time step $\Delta t=10^{-6}$. } 
\label{fig:2D_Jt_kap_H10}
\end{figure*}

\subsection{Mean Escape Time}\label{grooves_mean}

In this section, we compute the mean escape time by analyzing the asymptotic
behavior of $\tilde{J}_{\rm{ab}}(s|r,z)$ as $s\to 0$. In the limit $s\to 0$, one has 
\begin{equation}
\alpha_n^{(s)} \approx \begin{cases} \sqrt{s \kappa'_a/k'_d} \quad\quad\quad\quad\hspace{0.9ex} (n=0) ,\\ 
\pi n + \kappa'_a s/(k'_d \pi n) \quad (n > 0), \end{cases}
\end{equation}
where we used $\alpha_n^{(0)} = \pi n$.  We deduce then 
\begin{equation}
c_n^{(s)} \approx \begin{cases} 1 + \frac{\kappa'_a}{6k'_d} s \quad\quad\quad\quad\quad\quad\hspace{1ex} (n=0) ,\\  2s(-1)^n \kappa'_a/(k'_d [\pi n]^2) \quad (n > 0). \end{cases}
\end{equation}
Similarly, we deduce
\begin{equation}
\frac{g_0^{(s)}(z)}{g_0^{(s)}(H)} \approx 1 - \biggl(\frac{\kappa_a}{k_d} \frac{H-z}{L} + \biggl[\frac{L^2}{D} + \frac{\kappa'_a}{k'_d}\biggr]
\frac{H^2-z^2}{2L^2}\biggr) s, 
\end{equation}
and 
\begin{equation}
\frac{g_n^{(s)}(z)}{g_n^{(s)}(H)} \approx \frac{\cosh(\pi n z/L)}{\cosh(\pi n H/L)} \,.
\end{equation}
Substituting these expressions in Eq. (\ref{eq:Jtilde_eq_2D}), we get
$\tilde{J}_{\rm{ab}}(s|x,z) = 1 - s \langle T(x,z) \rangle + O(s^2)$, where the mean escape time is found to be
\begin{align} 
\langle T(x,z) \rangle & = \frac{H^2-z^2}{2D} + \frac{k_a}{k_d} \frac{H-z}{D} \\ \nonumber
& + \frac{k'_a L}{k'_d D} \biggl(\frac{x^2}{2L^2} - \frac16  + \frac{H^2-z^2}{2L^2} \\
& - 2\sum\limits_{n=1}^\infty \frac{(-1)^n}{\pi^2 n^2} \cos(\pi nx/L) \frac{\cosh(\pi nz/L)}{\cosh(\pi nH/L)} \biggr).  \nonumber
\end{align}
The average over the cross-section at height $z$ yields
\begin{align} \nonumber
\langle \overline{T}(z) \rangle & = \frac{1}{L} \int\limits_{0}^L dx \, T(x,z) \\
& =  \frac{k_a (H-z)}{k_d D} + \biggl(1 + \frac{k'_a}{k'_d L}\biggr) \frac{H^2-z^2}{2D} .
\end{align}
Further averaging over $z$ we obtain 
\begin{align}\label{mean_uniform_3}  
\langle \mathcal{T}_{u}\rangle  &= \frac{1}{H} \int\limits_{0}^H dz \langle \overline{T}(z) \rangle \\
& = \frac{H^2}{3 D} \left( 1 + \frac{k_a}{k_d}\frac{3}{2H}  + \frac{k'_a}{k'_d}\frac{1}{L}  \right) , \nonumber
\end{align}
where the subscript `$u$' denotes a uniform distribution of the initial position.

If one uses the two-state switching diffusion approximation instead of the exact solution (see Sec. \ref{two_state_diffsion}), a comparison with Eq. (\ref{eq:Tmean_switch}) for $k_a = 0$ suggests
that $k_{12} = k'_a/L$.

\subsection{Solution in time domain}\label{grooves_time_domain}

The solution in time domain can be found via the residue theorem.  The
computation is similar to the case of the capped cylinder in Sec. \ref{holes_time_domain}; here, we
differentiate Eq. (\ref{eq:alpha_n_eq_2D}) to get
\begin{align}  \label{eq:alp_dalp_2D}
\alpha_n^{(s)} \frac{d\alpha_n^{(s)}}{ds} &=  \\ 
& \frac{\kappa'_a k'_d}{(k'_d + s)^2} \biggl(\frac{\tan(\alpha_n^{(s)})}{\alpha_n^{(s)}} 
+ 1 + \tan^2(\alpha_n^{(s)})\biggr)^{-1} .\nonumber
\end{align}
Therefore, we obtain
\begin{align} \label{eq:Jt_2D_exact}
& J_{\rm{ab}}(t|x,z)  = \sum\limits_{n,m=0}^\infty c_{n,m} e^{-Dt\lambda_{n,m}/L^2} \cos(\alpha_{n,m} x/L) \times \nonumber \\  
&  \biggl(\beta_{n,m} \cos(\beta_{n,m} z/L) - \frac{\kappa_a \lambda_{n,m}}{\kappa_d - \lambda_{n,m}} \sin(\beta_{n,m} z/L)\biggr), 
\end{align}
where $\alpha_{n,m} = \alpha_n^{(s_{n,m})}$, $\lambda_{n,m} =
\alpha_{n,m}^2 + \beta_{n,m}^2 = - s_{n,m} L^2/D$,
\begin{align} 
c_{n,m} & = \frac{i c_n^{(s_{n,m})}}{\frac{dg_n^{(s)}(H)}{ds}\bigr|_{s=s_{n,m}}} \\
& = \frac{i}{\frac{dg_n^{(s)}(H)}{ds} \bigr|_{s=s_{n,m}}}
\frac{2\sin(\alpha_{n,m})/\alpha_{n,m}}{1 + \frac{\sin(2\alpha_{n,m})}{2\alpha_{n,m}}} , \nonumber
\end{align}
and $\frac{dg_n(H)}{ds}|_{s=s_{n,m}}$ is given by
Eq. (\ref{eq:dgnH_ds}), in which $\alpha_n^{(s)}
\frac{d\alpha_n^{(s)}}{ds}$ is substituted from
Eq. (\ref{eq:alp_dalp_2D}).

Figure \ref{fig:2D_Jt_kap_H10} illustrates the behavior of the PDF $J_{\rm{ab}}(t|x,z)$ for $H = 10$.  For small adsorption rates $\kappa'_a$ (panels a and b), the two-state switching diffusion model yields an excellent approximation.  In contrast, when $\kappa'_a = 10$ (panel c), the two-state model accurately describes the long-time behavior but fails at short times.  Similarly, when $H = 1$, an approximation by the two-state model is less accurate (figure not shown). 


\begin{figure*}[t]
\begin{center}
\includegraphics[width=\linewidth]{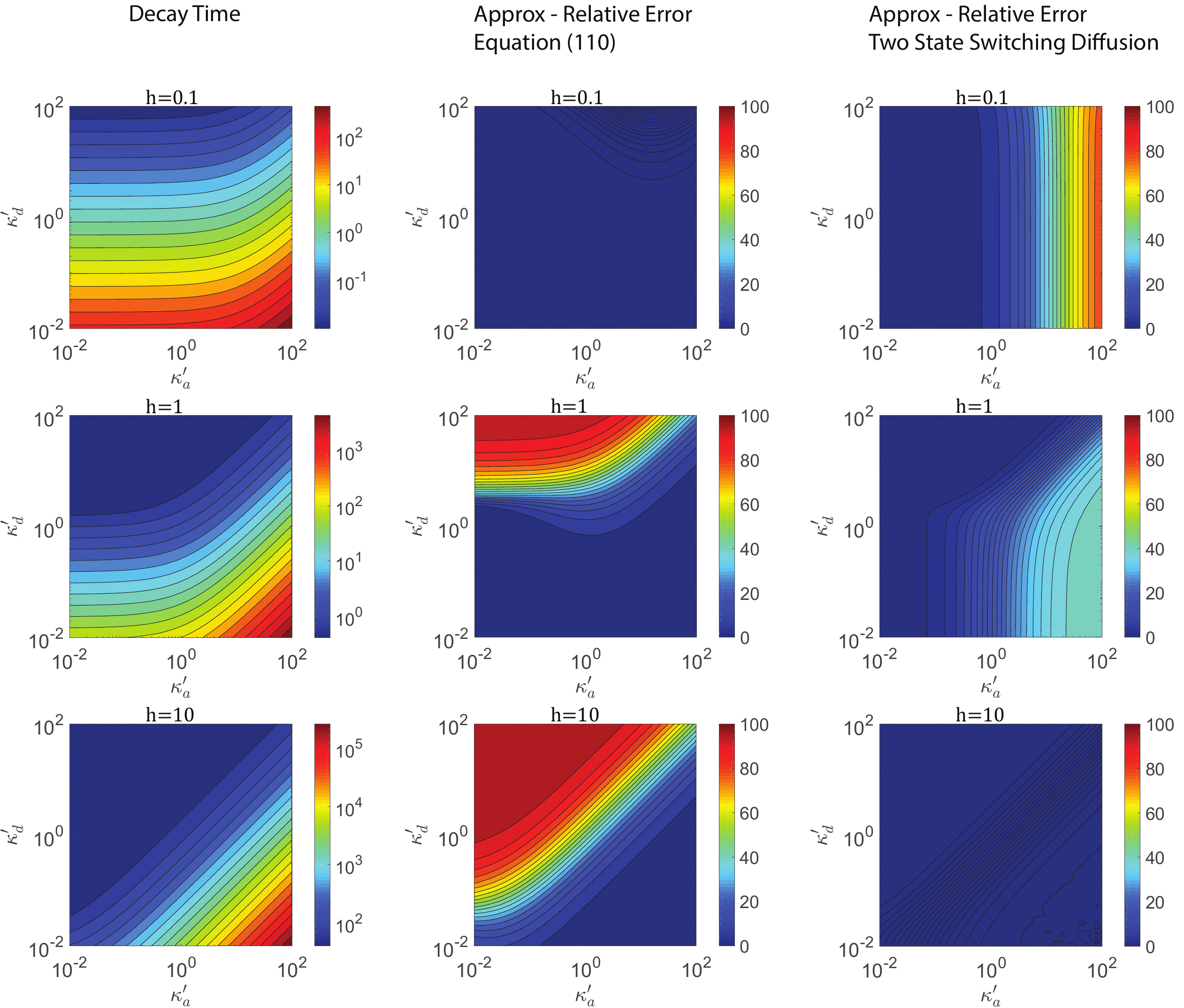}
\end{center}
\caption{\textbf{Left Column.} The decay time $T$ for the escape from a groove with $h=H/L=0.1$ (top), $h = 1$
(middle), and $h = 10$ (bottom), for $k_a = 0$, $L=1$
and $D = 1$. Since $k_a=0$ we have $\beta_0=\pi/(2h)$. We attain $\bar{\alpha}_0$ by numerically finding the roots of Eq. (\ref{eq:auxil3}). We then use $\lambda_{0,0} = \beta_0^2 - \bar{\alpha}_0^2$ and $T=L^2/(D\lambda_{0,0})$. \textbf{Center Column.} The relative error (in per cents) of the approximation for $T$ given by Eq. (\ref{eq:rec_approx}). \textbf{Right Column.} The relative error (in per cents) of the two-state switching diffusion approximation for $T$.}
\label{fig:2D_Tdecay}
\end{figure*}

\subsection{Poles}\label{grooves_poles}
The poles of $\tilde{J}_{\rm{ab}}(s|x,z)$ are determined by zeros of
$g_n^{(s)}(H)$, as previously (see Sec. \ref{holes_poles}).  We use the former notations:
$\alpha_n^{(s)} = \alpha$, $\hat{\alpha}_n^{(s)} = i\beta$, and
$\alpha^2 + \beta^2 = \lambda = - s L^2/D$.  In this case,
Eq. (\ref{eq:alpha_n_eq_2D}) reads
\begin{equation}  \label{eq:system1_2D}
\frac{\cos(\alpha)}{\alpha \sin(\alpha)} = \frac{1 - \kappa'_d/(\alpha^2 + \beta^2)}{\kappa'_a} \,.
\end{equation}

We focus on the case $k_a = 0$, for which $\beta_n h = \pi/2 + \pi n$ with $h = H/L$.
For any fixed $\beta_n$, the left-hand side of
Eq. (\ref{eq:system1_2D}) decreases piecewise monotonously on the
intervals $(\pi m, \pi(m+1))$, while the right-hand side is a
monotonously increasing function of $\alpha$.  As a consequence, there
is a single solution on each interval $(\pi m,\pi(m+1))$ that we
denote as $\alpha_{n,m+1}$, for $m = 0,1,2,\ldots$.  In addition,
there is a purely imaginary solution, which can be found by setting
$\alpha = -i\bar{\alpha}$, with $\bar{\alpha}$ satisfying
\begin{equation}  \label{eq:auxil3}
\frac{\cosh(\bar{\alpha})}{\bar{\alpha} \sinh(\bar{\alpha})} = \frac{\kappa'_d/(\beta_n^2 - \bar{\alpha}^2) - 1}{\kappa'_a}  \,.
\end{equation}
As
$\bar{\alpha}$ goes from $0$ to $+\infty$, the left-hand side monotonously decreases from $+\infty$ to $0$,
whereas the right-hand side for any fixed $\beta_n$ increases
monotonously on $(0,\beta_n)$ from $(\kappa'_d/\beta_n^2 -
1)/\kappa'_a$ to $+\infty$, and on $(\beta_n,+\infty)$ from $-\infty$
to $-1/\kappa'_a$.  As a consequence, there exists only one solution
on the interval $(0,\beta_n)$ that we denote $\bar{\alpha}_n$.  This
solution determines $\alpha_{n,0} = - i\bar{\alpha}_n$ that
contributes to the list of poles.

\subsection{Decay time}\label{sec:grooves_decay}

The general discussion in Sec. \ref{sec:holes_decay} is valid here. Let us find the approximation for the decay time $T$ in the limit $k_a=0$ and $h =H/L \ll 1$ such that $\beta_0 = \pi/(2h) \gg 1$. A solution of Eq. (\ref{eq:auxil3}) can be searched (and then validated with simulations) by setting $\bar{\alpha}_0 = \pi/(2h) - \epsilon$ with $\epsilon \ll 1$.  Substituting this expression and expanding in the leading order to $\epsilon$, we get
\begin{equation}
\epsilon \approx \frac{\kappa'_d}{\frac{\pi}{h} + 2\kappa'_a \ctanh(\pi/(2h))} ,
\end{equation}
from which $\lambda_{0,0} = \beta_0^2 - \bar{\alpha}_0^2 \approx \pi
\epsilon/h$ and thus
\begin{align} \label{eq:rec_approx}
T & = \frac{L^2}{D\lambda_{0,0}} \approx \frac{L^2}{D} \frac{1 + \frac{2}{\pi} \kappa'_a h\, \ctanh(\pi/(2h))}{\kappa'_d}  \\  
& = \frac{1}{k'_d} + \frac{\frac{2}{\pi} k'_a (H/D) \ctanh(\pi L/(2H))}{k'_d} \nonumber\\ 
&\approx \frac{1}{k'_d} \left( 1 + \frac{2}{\pi} \frac{k'_a H}{D}\right)
,\nonumber
\end{align}
where we used $\ctanh(\pi L/(2H))\approx 1$ when $h \ll 1$. In this limit the decay time does not depend on $L$. Note that the decay time in this limit is the same as in the previous two examples, see Eqs. (\ref{eq:decay}) and (\ref{eq:decay_2}).  

For this example let us also consider the opposite limit $h \gg 1$. In fact, we have already derived in Sec. \ref{two_state_diffsion} the two-state switching diffusion approximation for this case. The decay time of this approximation is determined by $\gamma_-$ with the lowest eigenvalue $\Lambda_0=(\pi/2H)^2$, according to Eq. (\ref{gamma_def}).  At the end of Sec. \ref{grooves_mean} we have already seen that for this example $k_{12} = k'_a/L$ and $k_{21} = k'_d$. Finally, we have $T_{\rm sd} =L^2/(D\gamma_-)$.  

The  behavior of the decay time and the validity of the approximation in Eq. (\ref{eq:rec_approx}) and the two-state switching diffusion approximation $T_{\rm sd}$ are explored in Fig.  \ref{fig:2D_Tdecay}, where we plot the actual decay time as a function of $\kappa'_a$ and $\kappa'_d$, and the relative errors that are obtained by using the approximations. While it can be seen that both approximations work very well in their ranges of validity ($h\ll1$ for Eq. (\ref{eq:rec_approx}) and $h\gg1$ for two-state diffusion), we observe that the two-state diffusion approximation also provides fair results for grooves of intermediate depth ($h=1)$.


\section{Discussion}\label{sec:discussion}

Since the early works of Langmuir in the beginning of the 20th century, and to this day, the vast majority of theoretical works on adsorption dynamics have dealt with \textit{flat} surfaces immersed in an infinite bulk of adsorbates  \cite{langmuir1918adsorption,brunauer1938adsorption,ward1946time,sutherland1952kinetics,delahay1957adsorption,hansen1961diffusion,baret1968kinetics,miller1981solution,mccoy1983analytical,adamczyk1987nonequilibrium,miller1991adsorption,chang1995adsorption,liggieri1996diffusion,diamant1996kinetics,liu2009diffusion,foo2010insights,miura2015diffusion,miller2017dynamic,noskov2020adsorption} (which is effectively a one-dimensional setting), or with smooth curved surfaces \cite{mysels1982diffusion,frisch1983diffusion,adamczyk1987adsorption,reva2021first,prustel2012exact,kim1999exact,grebenkov2019reversible,scher2022microscopic,grebenkov2023diffusion}. 
In this context, two main models are usually considered: Linear kinetics and Langmuir kinetics. While the latter accounts for saturation of the surface under high adsorbate concentration, it is not linear and does not admit an analytical solution.

Given a flat surface located at $L$ and a concentration profile $c(x,t)$, the surface concentration $\Gamma(t)$ under linear adsorption kinetics follows \cite{scher2022microscopic}  
\begin{equation} \label{Henry}
 \frac{d \Gamma(t)}{d t} = k_{a} c(L,t) - k_{d} \Gamma(t).
\end{equation}
At equilibrium, i.e., when $ \frac{d \Gamma(t)}{d t} = 0 $, one gets the Henry isotherm $\Gamma(t) = K c(L,t)$, with the equilibrium constant $K = k_{a}/k_{d}$. Trivially, the desorption rate $k_d$ is inversely proportional to the mean time spent being adsorbed to the surface: $ k_d = \langle \mathcal{T} \rangle^{-1}$. But what happens when the surface is textured? On the microscopic scale, the desorption rate is still determined by $k_d$. Yet, on a scale comparable to that of the surface roughness, multiple events of adsorption and desorption can give rise to a completely different {\it effective} (or macroscopic) desorption rate. In this work, we employed our theoretical approach presented in Ref. \cite{scher2023escape} to determine this effective desorption rate.

For this purpose, we adopted a single-particle perspective and calculated the PDF and the mean of the escape time from textured surfaces of three different common topographies: holes, pillars and grooves. Such solutions are valuable when studying the adsorption-desorption dynamics of surfaces. In particular, we obtained the mean escape time for the three surface topographies in a common experimental setting where the initial position of the particle is uniformly distributed inside the surface cavities. Let us rewrite these equations (\ref{mean_uniform}, \ref{mean_uniform_2}, \ref{mean_uniform_3}) here
\begin{subequations} 
\begin{align}
 & \langle {\mathcal{T}_{holes}}\rangle  = \frac{H^2}{3D} \left(1 + K \frac{3}{2H}  + K' \frac{2}{L}  \right), \label{mean_uniform_compare_holes} \\
 & \langle {\mathcal{T}_{pillars}}\rangle  = \frac{H^2}{3D} \left( 1+ K  \frac{3}{2H} + K' \frac{\rho}{1-\rho^2} \frac{2}{L} \right), \label{mean_uniform_compare_pillars}\\
 & \langle {\mathcal{T}_{grooves}}\rangle  = \frac{H^2}{3 D} \left( 1 + K\frac{3}{2H}  + K'\frac{1}{L}  \right) \label{mean_uniform_compare_grooves},
\end{align}
\end{subequations}
where $K=k_a/k_d$ and $K'=k'_a/k'_d$ are the equilibrium constants of the bottom and lateral surfaces respectively, and $\rho = l/L$.  

For all the considered examples, the first term in the parenthesis corresponds to the mean escape time in the absence of adsorption (reflecting surfaces): $\langle \mathcal{T}_{ref}\rangle=H^2/(3D)$. The second and third terms correspond to the mean time spent adsorbed to the bottom surface and to the lateral surface, respectively.  We can easily quantify how the introduction of stickiness affects the mean escape time. For example,  dividing Eq. (\ref{mean_uniform_compare_holes}) by $\langle \mathcal{T}_{ref}\rangle$ we obtain
\begin{equation} \label{mean_free}
\frac{\langle \mathcal{T}_{holes}\rangle}{\langle \mathcal{T}_{ref}\rangle}  = 1 + K \frac{3}{2H}  + K' \frac{2}{L}  \,.
\end{equation}
It is apparent that there are two contributions to the deviation of the mean escape time from its non-sticky benchmark. The contribution from the bottom surface scales like $K/H$ and the contribution from the lateral surface scales like $K'/L$, where $L$ and $H$ are the characteristic lengths in the direction of the axes. As first mentioned in Ref. \cite{scher2023escape} this scaling seems to be universal, and the geometry of the domain determines the effective length $\xi$. It can be easily verified to hold in Eqs. (\ref{mean_uniform_compare_pillars}) and (\ref{mean_uniform_compare_grooves}). 

In this work we considered surfaces with canonical topographies. However, the approach we employed is general and we expect that similar behavior will also be found for other geometries with perpendicular sticky surfaces, up to the geometrical pre-factor $\xi$. For a general (even rugged) sticky surface, the expected form of the mean escape time is 
\begin{equation} \label{mean_uniform_norm}
\langle \mathcal{T}_{u}\rangle  = \langle \mathcal{T}_{ref}\rangle \left( 1 + \sum_n \frac{K_n}{\xi_n} \right),
\end{equation}
where $\xi_n$ is the effective length scale of the $n$-th surface element, and $K_n = k^{(n)}_a/k^{(n)}_d$ is the equilibrium constant with $k^{(n)}_a$ and $k^{(n)}_d$ standing for the adsorption and desorption rates.

Finally, the mean escape time is inversely proportional to the effective desorption rate from the textured surface
\begin{equation} \label{effective_desorption}
 k^{eff}_{d} = \langle \mathcal{T}_{u}\rangle^{-1}  = \left[\langle \mathcal{T}_{ref}\rangle \left( 1 + \sum_n \frac{K_n}{\xi_n} \right)\right]^{-1}.
\end{equation}
Note that Eq. (\ref{effective_desorption}) represents an effective macroscopic description of the system, and that it does not imply exponential escape times from the surface. Importantly, $k^{eff}_{d}$ can be orders of magnitude smaller than the characteristic (microscopic) desorption rates in the system, as it vanishes with  lateral confining length scales. Equation (\ref{effective_desorption}) can guide those who wish to predict and control desorption from textured adsorbing surfaces.

Overall, our study can also be considered as the fundamental step towards coarse-graining the microscopic adsorption-desorption kinetics and thus building macroscopic models of diffusive dynamics near textured surfaces. In fact, former works on boundary homogenization (see \cite{grebenkov2022diffusion,grebenkov2023pillar} and references therein)
provided efficient tools for estimating the macroscopic {\it adsorption} constant for textured surfaces. This work complements the former results by quantifying the desorption step and thus opening a way to describe adsorption-desorption kinetics of textured surfaces. 


\textit{Acknowledgments.}---We thank Samyuktha Ganesh for help with the graphical design. This project has received funding from the European Research Council (ERC) under the European Union’s Horizon 2020 research and innovation program (Grant agreement No. 947731). D.G. acknowledges the Alexander von Humboldt Foundation for support within a Bessel Prize award.



\begin{thebibliography}{10}


\bibitem{lord2010influence}
M.S. Lord, M. Foss, F. Besenbacher (2010). Influence of nanoscale surface topography on protein adsorption and cellular response. \textit{Nano Today}, 5(1), 66-78.

\bibitem{flemming1999effects}
R.G. Flemming, C.J. Murphy, G.A. Abrams, S.L. Goodman, P.F. Nealey (1999). Effects of synthetic micro-and nano-structured surfaces on cell behavior.\textit{ Biomaterials}, 20(6), 573-588.

\bibitem{alhmoud2021maceing}
H. Alhmoud, D. Brodoceanu, D. R. Elnathan, T. Kraus, N.H. Voelcker (2021). A MACEing silicon: Towards single-step etching of defined porous nanostructures for biomedicine. \textit{Progress in materials science}, 116, 100636.

\bibitem{hochbaum2005controlled}
A.I. Hochbaum, R. Fan, R. He, P. Yang (2005). Controlled growth of Si nanowire arrays for device integration. \textit{Nano letters}, 5(3), 457-460.

\bibitem{elnathan2014engineering}
R. Elnathan, M. Kwiat, F. Patolsky, N.H. Voelcker (2014). Engineering vertically aligned semiconductor nanowire arrays for applications in the life sciences. \textit{Nano Today}, 9(2), 172-196.

\bibitem{sun2014large}
L. Sun, Y. Fan, X. Wang, R.A. Susantyoko, Q. Zhang (2014). Large scale low cost fabrication of diameter controllable silicon nanowire arrays. \textit{Nanotechnology}, 25(25), 255302.

\bibitem{checco2014collapse}
A. Checco, B.M. Ocko, A. Rahman, C.T. Black, M. Tasinkevych, A. Giacomello, S. Dietrich (2014). Collapse and reversibility of the superhydrophobic state on nanotextured surfaces. \textit{Physical Review Letters}, 112(21), 216101.

\bibitem{chen2022nanoforest}
G. Chen, R. Guan, M. Shi, X. Dai, H. Li, N. Zhou,  D. Chen, H. Mao (2022). A nanoforest-based humidity sensor for respiration monitoring. \textit{Microsystems \& Nanoengineering}, 8(1), 44.


\bibitem{richert2010adsorption}
L. Richert, F. Variola, F. Rosei, J.D. Wuest, A. Nanci (2010). Adsorption of proteins on nanoporous Ti surfaces. \textit{Surface Science}, 604(17-18), 1445-1451.

\bibitem{wilkinson2011biomimetic}
A. Wilkinson, R. Hewitt, L.E. McNamara, D. McCloy, R.D. Meek, M.J. Dalby (2011). Biomimetic microtopography to enhance osteogenesis in vitro. \textit{Acta biomaterialia}, 7(7), 2919-2925.

\bibitem{tsai2009fibronectin}
W.B. Tsai, Y.C. Ting, J.Y. Yang, J.Y. Lai, H.L. Liu (2009). Fibronectin modulates the morphology of osteoblast-like cells (MG-63) on nano-grooved substrates. \textit{Journal of Materials Science: Materials in Medicine}, 20, 1367-1378.

\bibitem{de2015effect}
A.C. De Luca, M. Zink, A. Weidt, S.G. Mayr, A.E. Markaki (2015). Effect of microgrooved surface topography on osteoblast maturation and protein adsorption. \textit{Journal of Biomedical Materials Research Part A}, 103(8), 2689-2700.

\bibitem{van2008manufacturing}
F.C.M.J.M. Van Delft, F. C. Van Den Heuvel, W.A. Loesberg, J. Te Riet, P. Schön, C.G. Figdor, S. Speller, J. J. W. A. Van Loon, X. F. Walboomers, J.A. Jansen (2008). Manufacturing substrate nano-grooves for studying cell alignment and adhesion. \textit{Microelectronic Engineering}, 85(5-6), 1362-1366.

\bibitem{coppens1999effect}
M.O. Coppens (1999). The effect of fractal surface roughness on diffusion and reaction in porous catalysts–from fundamentals to practical applications. \textit{Catalysis Today}, 53(2), 225-243.

\bibitem{ben2000diffusion}
D. ben-Avraham, S. Havlin, \textit{Diffusion and Reactions in Fractals and
Disordered Systems} (Cambridge University Press, 2010)

\bibitem{filoche2005deactivation}
M. Filoche, B. Sapoval, J.S. Andrade Jr (2005). Deactivation dynamics of rough catalytic surfaces. \textit{AIChE journal}, 51(3), 998-1008.

\bibitem{filoche2008passivation}
M. Filoche, D.S. Grebenkov, J.S. Andrade Jr, B. Sapoval (2008). Passivation of irregular surfaces accessed by diffusion. \textit{Proceedings of the National Academy of Sciences}, 105(22), 7636-7640.

\bibitem{chen2009effect}
M. Chen, C. Wu, D. Song, W. Dong, K. Li (2009). Effect of grooves on adsorption of RGD tripeptide onto rutile Ti$_2$ (110) surface. \textit{Journal of Materials Science: Materials in Medicine}, 20, 1831-1838.

\bibitem{roach2007modern}
P. Roach, D. Eglin, K. Rohde, C.C. Perry (2007). Modern biomaterials: a review—bulk properties and implications of surface modifications. \textit{Journal of Materials Science: Materials in Medicine}, 18, 1263-1277.

\bibitem{wilson2005mediation}
C.J. Wilson, R.E. Clegg, D.I. Leavesley, M.J. Pearcy  (2005). Mediation of biomaterial–cell interactions by adsorbed proteins: a review. \textit{Tissue engineering}, 11(1-2), 1-18.

\bibitem{peng2013micropatterned}
W. Peng, Z. Qiao, Q. Zhang, , X. Cao, X. Chen, H. Dong, J. Liao, C. Ning (2013). Micropatterned TiO$_{2}$ nanotubes: fabrication, characterization and in vitro protein/cell responses. \textit{Journal of Materials Chemistry B}, 1(28), 3506-3512.

\bibitem{borberg2019light}
E. Borberg, M. Zverzhinetsky, A. Krivitsky, A. Kosloff, O. Heifler, G. Degabli, H. Peretz-Soroka, R. Satchi-Fainaro, L. Burstein, S. Reuveni, H. Diamant, V. Krivitsky and F. Patolsky (2019). Light-controlled selective collection-and-release of biomolecules by an on-chip nanostructured device. \textit{Nano letters}, 19(9), 5868-5878.

\bibitem{borberg2021depletion}
E. Borberg, S. Pashko, V. Koren, L. Burstein, F. Patolsky (2021). Depletion of highly abundant protein species from biosamples by the use of a branched silicon nanopillar on-chip platform. \textit{Analytical Chemistry}, 93(43), 14527-14536.

\bibitem{krivitsky2012si}
V. Krivitsky, L.C. Hsiung, A. Lichtenstein, B. Brudnik, R. Kantaev, R. Elnathan,  A. Pevzner, A. Khatchtourints, F. Patolsky (2012). Si nanowires forest-based on-chip biomolecular filtering, separation and preconcentration devices: nanowires do it all. \textit{Nano letters}, 12(9), 4748-4756.

\bibitem{scher2023escape}
Y. Scher, S. Reuveni, D.S. Grebenkov (2023). Escape of a Sticky Particle. 
\textit{Physical Review Research}, 5(4), 043196.



\bibitem{karger1985nmr}
J. Kärger (1985). NMR self-diffusion studies in heterogeneous systems. \textit{Advances in Colloid and Interface Science}, 23, 129-148.


\bibitem{Godec17}		A. Godec and R. Metzler (2017).
				First passage time statistics for two-channel diffusion,
\textit{Journal of Physics A: Mathematical and Theoretical}, 50(8), 084001.

\bibitem{Grebenkov19a}		D. S. Grebenkov (2019).
				A unifying approach to first-passage time distributions in diffusing diffusivity and switching diffusion models,
				\textit{Journal of Physics A: Mathematical and Theoretical}, 52(17), 174001. 

\bibitem{redner2001guide}
S. Redner, \textit{A guide to first-passage processes}. (Cambridge
University Press, Cambridge, England, 2001).

\bibitem{grebenkov2022diffusion}
D.S. Grebenkov, A.T. Skvortsov (2022). Diffusion toward a nanoforest of absorbing pillars. \textit{The Journal of Chemical Physics}, 157(24), 244102.

\bibitem{grebenkov2023pillar}
D. S. Grebenkov, A. T. Skvortsov (2023). Survival in a nanoforest of absorbing pillars. \textit{Journal of Physics A: Mathematical and Theoretical}, 56(16), 165002.

\bibitem{langmuir1918adsorption}
I. Langmuir (1918). The adsorption of gases on plane surfaces of glass, mica and platinum. \textit{Journal of the American Chemical society}, 40(9), 1361-1403.

\bibitem{brunauer1938adsorption}
S. Brunauer, P.H. Emmett, E. Teller (1938). Adsorption of gases in multimolecular layers. \textit{Journal of the American chemical society}, 60(2), 309-319.

\bibitem{ward1946time}
A.F.H. Ward, L. Tordai (1946). Time‐dependence of boundary tensions of solutions I. The role of diffusion in time‐effects. \textit{The Journal of Chemical Physics}, 14(7), 453-461.

\bibitem{sutherland1952kinetics}
K.L. Sutherland. (1952). The kinetics of adsorption at liquid surfaces. \textit{Australian Journal of Chemistry}, 5(4), 683-696.

\bibitem{delahay1957adsorption}
P. Delahay, I. Trachtenberg (1957). Adsorption kinetics and electrode processes. \textit{Journal of the American Chemical Society}, 79(10), 2355-2362.

\bibitem{hansen1961diffusion}
R.S. Hansen (1961). Diffusion and the kinetics of adsorption of aliphatic acids and alcohols at the water-air interface. \textit{Journal of Colloid Science}, 16(6), 549-560.

\bibitem{baret1968kinetics}
J.F. Baret (1968). Kinetics of adsorption from a solution. Role of the diffusion and of the adsorption-desorption antagonism. \textit{The Journal of Physical Chemistry}, 72(8), 2755-2758.

\bibitem{miller1981solution}
R. Miller (1981). On the solution of diffusion controlled adsorption kinetics for any adsorption isotherms. \textit{Colloid and polymer science}, 259(3), 375-381.

\bibitem{mccoy1983analytical}
B.J. McCoy (1983). Analytical solutions for diffusion-controlled adsorption kinetics with non-linear adsorption isotherms. \textit{Colloid and polymer science}, 261(6), 535-539.

\bibitem{adamczyk1987nonequilibrium}
Z. Adamczyk (1987). Nonequilibrium surface tension for mixed adsorption kinetics. \textit{Journal of colloid and interface science}, 120(2), 477-485.

\bibitem{miller1991adsorption}
R. Miller, G. Kretzschmar (1991). Adsorption kinetics of surfactants at fluid interfaces. \textit{Advances in colloid and interface science}, 37(1-2), 97-121.

\bibitem{chang1995adsorption}
C.H. Chang, E.I. Franses (1995). Adsorption dynamics of surfactants at the air/water interface: a critical review of mathematical models, data, and mechanisms. \textit{Colloids and Surfaces A: Physicochemical and Engineering Aspects}, 100, 1-45.

\bibitem{liggieri1996diffusion}
L. Liggieri, F. Ravera, A. Passerone (1996). A diffusion-based approach to mixed adsorption kinetics. \textit{Colloids and surfaces A: physicochemical and engineering aspects}, 114, 351-359.

\bibitem{diamant1996kinetics}
H. Diamant, D. Andelman (1996). Kinetics of surfactant adsorption at fluid-fluid interfaces. \textit{The Journal of Physical Chemistry}, 100(32), 13732-13742.

\bibitem{liu2009diffusion}
J. Liu, P. Li, C. Li, Y. Wang (2009). Diffusion-controlled adsorption kinetics of aqueous micellar solution at air/solution interface. \textit{Colloid and Polymer Science}, 287(9), 1083-1088.

\bibitem{foo2010insights}
K.Y. Foo, B.H. Hameed (2010). Insights into the modeling of adsorption isotherm systems. \textit{Chemical engineering journal}, 156(1), 2-10.

\bibitem{miura2015diffusion}
T. Miura, K. Seki (2015). Diffusion influenced adsorption kinetics. \textit{The Journal of Physical Chemistry B}, 119(34), 10954-10961.

\bibitem{miller2017dynamic}
R. Miller, E.V. Aksenenko, V.B. Fainerman (2017). Dynamic interfacial tension of surfactant solutions. \textit{Advances in colloid and interface science}, 247, 115-129.

\bibitem{noskov2020adsorption}
B.A. Noskov, A.G. Bykov, G. Gochev, S.Y. Lin, G. Loglio, R. Miller, O.Y. Milyaeva (2020). Adsorption layer formation in dispersions of protein aggregates. \textit{Advances in colloid and interface science}, 276, 102086.

\bibitem{reva2021first}
M. Reva, D.A. DiGregorio, D.S. Grebenkov (2021). A first-passage approach to diffusion-influenced reversible binding and its insights into nanoscale signaling at the presynapse. \textit{Scientific reports}, 11(1), 1-17.


\bibitem{prustel2012exact}
T. Prüstel, M. Meier-Schellersheim (2012). Exact Green's function of the reversible diffusion-influenced reaction for an isolated pair in two dimensions. \textit{The Journal of chemical physics}, 137(5), 054104.

\bibitem{grebenkov2019reversible}
D.S. Grebenkov (2019). Reversible reactions controlled by surface diffusion on a sphere. \textit{The Journal of chemical physics}, 151(15), 154103.

\bibitem{kim1999exact}
H. Kim, K.J. Shin (1999). Exact solution of the reversible diffusion-influenced reaction for an isolated pair in three dimensions. \textit{Physical review letters}, 82(7), 1578.

\bibitem{mysels1982diffusion}
K.J. Mysels (1982). Diffusion-controlled adsorption kinetics. General solution and some applications. \textit{The Journal of Physical Chemistry}, 86(23), 4648-4651.

\bibitem{frisch1983diffusion}
H.J. Frisch, K.J. Mysels (1983). Diffusion-controlled adsorption. Concentration kinetics, ideal isotherms, and some applications. \textit{The Journal of Physical Chemistry}, 87(20), 3988-3990.

\bibitem{adamczyk1987adsorption}
Z. Adamczyk, J. Petlicki (1987). Adsorption and desorption kinetics of molecules and colloidal particles. \textit{Journal of colloid and interface science}, 118(1), 20-49.

\bibitem{grebenkov2023diffusion}
D.S. Grebenkov (2023). Diffusion-controlled reactions with non-Markovian binding/unbinding kinetics. \textit{The Journal of Chemical Physics}, 158(21), 214111.

\bibitem{scher2022microscopic}
Y. Scher, O. Lauber Bonomo, A. Pal, S. Reuveni (2023). Microscopic Theory of Adsorption Kinetics. \textit{The Journal of Chemical Physics}, 158, 094107.



\end{thebibliography}
\end{document}